\documentclass[preprint,pre,aps]{revtex4-2}
\usepackage[pdftex]{graphicx} 
\usepackage{xcolor}

\begin{document}
 \title{Spontaneous pattern formation in monolayers of binary mixtures with competing interactions }
 \author{ O. Patsahan}
\affiliation{Institute for Condensed Matter Physics of the National
	Academy of Sciences of Ukraine, 1 Svientsitskii St., 79011 Lviv,
	Ukraine}
	\author{ A. Meyra}
	\affiliation{1. Instituto de Física de Líquidos y Sistemas Bilógicos, UNLP-CONICET, 59-789, 1900 La Plata, Argentina \\
	 2. Depto. Ing. Mecánica 60 124, UTN-FRLP, 1900 La Plata, Argentina}
	\author{  A. Ciach\footnote{ the email address of the corresponding author:  alina.ciach@gmail.com }}
\affiliation{Institute of Physical Chemistry,  Polish Academy of Sciences, 01-224 Warszawa, Poland }
 \date{\today}
 \begin{abstract}
  	A model for a monolayer of two types of particles spontaneously forming ordered patterns
	is studied by a mesoscopic theory and by MC simulations.
  	We assume hard-cores of the same size for both components,  short-range attraction long-range repulsion between particles of the same species, and the cross-interaction of opposite sign. The model is inspired	by oppositely charged 
  	 particles or macromolecules with preferential solubility in different components of a solvent that is close to a miscibility critical point. 
  	We determine the phase diagram in the chemical potentials and in the concentration - density planes for a few fixed temperatures in the mean-field one-shell approximation. We find that the presence of the second component significantly enlarges the temperature range of stability of the ordered phases. We obtain three stable phases with periodic concentration: the lamellar L phase with alternating stripes of the two components for similar chemical potentials, and a hexagonal arrangement of the clusters of the minority component in the liquid of the majority component. The latter two phases, however, are stable only at relatively high temperatures. At lower temperatures, the L phase coexists with a disordered 
  one-component fluid or with very dilute gas with mixed components. 
   At still lower temperatures, the one-component phase coexisting with the L phase can be disordered or ordered, depending on the chemical potentials. The theoretical results are confirmed by MC simulations for selected thermodynamic states.
\end{abstract}

 \maketitle
 \section{introduction}
  Spontaneous pattern formation in soft matter or biological systems~\cite{seul:95:0,stradner:04:0,campbell:05:0,li:12:0,su:16:0} is often induced by competing interactions between particles or macromolecules. Due to the presence of electric charges, the particles or proteins repel each other at distances determined by the screening length, but  at shorter distances solvent-induced attraction can dominate.
The attraction can result in particular from the thermodynamic Casimir potential induced by critical concentration fluctuations in the solvent when it is close to the miscibility critical point~\cite{fisher:78:0,krech:99:0,hertlein:08:0}. An important example of a system that is close to the miscibility critical point is a multicomponent lipid bilayer 
in living organisms~\cite{veatch:07:0,machta:12:0}. Because of the fluctuation-induced Casimir interactions, membrane inclusions preferentially soluble in the same or in different membrane components attract or repel each other, respectively.
 The  electrostatic interactions between charged membrane proteins were typically disregarded,
 since the concentration of ions in their surroundings is large~\cite{eisberg:03:0}, and according to classical theories, the screening length should be very short.
 Recently, however, it was discovered that  in concentrated ionic solutions the screening length increases linearly with increasing density of ions and can become very large~\cite{smith:16:0,lee:17:0,ciach:23:0,safran:23:0}. Thus, the electrostatic interactions between the charged membrane inclusions may be important, and further studies are required to clarify their role.
 
While for particular membrane proteins the interactions 
resulting from the Casimir and the electrostatic forces 
are not known yet, in the case of charged selective particles in the critical mixture with small amount of dissolved ions,
 the interactions are known very well~\cite{pousaneh:12:0,hertlein:08:0}.
Parallel flat surfaces separated by a distance $L$ and
immersed in the near-critical mixture with ionic impurities interact with the potential
\begin{equation}
\label{V}
v(L)\approx A_C\exp(-L/\xi)+ A_{el}\exp(- L/\lambda_D),
\end{equation} 
where $\xi$ and $\lambda_D$
 are the correlation length of the concentration fluctuations in the solvent and the Debye screening length, respectively, and $A_C, A_{el}$ are the amplitudes. The amplitudes depend on $\xi$, $\lambda_D$ and on the charge and selectivity of the surfaces, and their theoretically predicted forms agree with experimental results obtained for small amount of salt present in the critical water-lutidine mixture~\cite{pousaneh:12:0,hertlein:08:0}.
 The first term in (\ref{V}) is attractive ($A_C<0$) for like adsorption preferences, and repulsive  ($A_C>0$) for opposite adsorption preferences.
The second term is repulsive ($A_{el}>0$) or attractive  ($A_{el}<0$) for like or opposite surface charges, respectively. 

For identical weakly charged and strongly selective  surfaces, the interactions can have a form of a short-range attraction long-range repulsion (SALR) when $\xi/\lambda_D <1$. On the other hand, if $\xi/\lambda_D <1$, oppositely charged hydrophilic and hydrophobic surfaces can repel and attract each other at short and large distances, respectively. Thus, the interaction between like particles and the cross-interaction are of opposite sign.   
Note that in view of the nonmonotonic dependence of the screening length on the density of ions,  the interactions can have similar features for small and large densities.
In the case of curved surfaces of spherical particles, the exponential terms in (\ref{V}) should be divided by the particle distance $L$, and the double Yukawa potential with opposite sign of the two terms is obtained.

The SALR potential is sometimes called 'a mermaid potential' because of an attractive head and a repulsive tail~\cite{royall:18:0}. Such interactions are interesting because they can lead to self-assembly into aggregates with the size determined by the shape of the potential~\cite{stradner:04:0,imperio:04:0,ciach:08:1,archer:08:0,zhuang:16:0}. 
The models with the SALR interactions were intensively studied by theoretical and simulation methods in three (3D) and two dimensions (2D)~\cite{imperio:04:0,Bartlett2005,imperio:06:0,archer:07:1,archer:08:0,ciach:08:1,pekalski:14:0,almarza:14:0,sweatman:14:0,lindquist:16:0,zhuang:16:0,zhuang:16:1,edelmann:16:0,pini:17:0,marlot:19:0,Liu2019,RuizFranco2021}, the latter case being suitable for the self-assembly in quasi 2D membranes, at fluid interfaces or at solid substrates. 
 The universal sequence of ordered patterns for  increasing density in 2D is the following: disordered gas - hexagonal pattern of clusters - stripes - hexagonal pattern of voids - disordered liquid~\cite{ciach:13:0,imperio:04:0,archer:08:0,pekalski:14:0,almarza:14:0}.  In the gas and the liquid, randomly distributed clusters and voids with well-defined size are present.
 If the interactions between the membrane inclusions were of the SALR form, then {\it clusters of well  defined size} would be present. Notably, clusters of membrane proteins are necessary for signaling and other important functions of life~\cite{li:12:0,su:16:0}.
 
It is natural to ask the question  how a presence of different particles or macromolecules  influences the self-assembly and pattern formation.  Mixtures of particles with competing interactions having spherical symmetry, however, were much less studied. Only recently several models for inhomogeneous binary mixtures
 were introduced and studied by theory and simulations~\cite{ciach:20:1,vonKonigslow_2013,FerreiroRangel2018,sweatman:21:0,scacchi:21:0,munao:22:0,costa:23:0}. As far as we know, 
 experimental studies of  self-assembly induced by the electrostatic and Casimir potentials were restricted to particles of the same type~\cite{shelke:13:0,nguyen:13:0,marino:21:0}.
As a result, fundamental questions such as the symmetries of the ordered phases and the sequence of patterns for varying chemical potentials and temperature  in mixtures with competing interactions remain open. 

 In this work we attempt to fill this gap by considering a model with the interactions inspired by the oppositely charged hydrophilic and hydrophobic particles   in near-critical solvent with $\xi/\lambda_D<1$.
 We assume the SALR interactions between like particles, and the cross interaction of the opposite sign. The model was introduced  in ref.\cite{ciach:20:1} and named 'two mermaids and a peacock' because of the mermaid potential between like particles, and a repulsive head and an attractive tail of the cross-interaction.
 We should stress that although the model is inspired by the oppositely charged particles or macromolecules with different adsorption preferences for the components of the critical solvent, it  is not restricted to such physical systems, and can be considered as a generic model for spontaneous pattern formation in binary mixtures  with competing interactions. Similar competing interactions may have different origin in different mixtures, and the self-assembled ordered patterns on the length scale of nanometers or micrometers may find numerous applications. 
 
A phase diagram for the two mermaids and a peacock  model was determined theoretically and by molecular dynamics (MD) simulations in Ref.\cite{patsahan:21:0} for equal chemical potentials of the two components in 3D. It was found that a gas coexists with a dense lamellar phase of alternating layers rich in the first and the second component. At low temperature $T$, the dense phase has the crystalline structure and the gas is very dilute. The crystal melts upon heating, but the alternating composition is preserved  in the liquid layers. The density difference between the gas and the lamellar phase decreases with increasing $T$. 
 
 In this work we determine the $(\mu_1,\mu_2,T)$ phase diagram for this mixture within the mesoscopic theory  in the mean-field approximation for a broad  range of the chemical potentials $\mu_1,\mu_2$ and temperature $T$ for a monolayer of the particles.
 The results are verified by Monte Carlo (MC) simulations for selected thermodynamic states.

 In sec.\ref{mt} we briefly summarize the mesoscopic theory in the mean-field (MF) approximation, and determine the boundary of stability of the disordered phase in sec.\ref{mf:bs}. The one-shell approximation and the method of obtaining the phase equilibria are described in sec.\ref{sec:one_shell}. In sec.\ref{sec:model} we introduce the interaction potential for which the theoretical and simulation
 results are obtained. The theoretical  results are described in sec.\ref{sec:theory_results}, and  in sec.\ref{MC} the representative patterns obtained in the MC simulations are presented. In sec.\ref{sec:discuss} we discuss and summarize the results.
  
 \section{mesoscopic theory for symmetrical binary mixtures} 
 \subsection{Mesoscopic grand-potential functional in mean-field approximation}
 \label{mt}
 The theory for mixtures with spontaneously formed inhomogeneities on the mesoscopic length scale was developed in Ref.\cite{ciach:11:2}. Here we briefly summarize it for binary symmetrical mixtures of particles having the same diameter $a$ that sets the length unit. 
For simplicity of calculations, we assume that the interaction between  particles of the same component is $u_{11}=u_{22}=u$, and depends only on the distance $r$ between the particles.
 In addition,
we assume that for $r>1$ (in $a$-units), $u(r)$  is attractive at short and repulsive at large distances (SALR or 'mermaid' potential~\cite{ciach:08:1,royall:18:0}).
 With such interactions, the particles in the one component system  self-assemble into various aggregates and can form regular patterns at sufficiently low temperature $T$~\cite{ciach:10:1,archer:08:0,zhuang:16:0,pini:17:0}.
 Finally, we assume that the cross-interaction is of opposite sign, and to simplify the calculations, we postulate that
  $u_{12}(r)=-u(r)$ for $r>1$. 
  
 For description of the ordering on the mesoscopic length scale, we consider volume fractions $\zeta_i({\bf r})$ with $i=1,2$  in the mesoscopic regions around ${\bf r}$. For given forms of $\zeta_i({\bf r})$, i.e. with neglected fluctuations on the mesoscopic length scale, the pair distribution function for our symmetrical mixture  is approximated by $g_{ij}=\theta(r-1)$, and the internal energy takes the  form
 \begin{equation}
 \label{U}
 U=\frac{1}{2}\int d{\bf r}_1\int d{\bf r} c({\bf r}_1)V(r)c({\bf r}_1+{\bf r})=\frac{1}{2}\int d{\bf k} \hat c({\bf k})\tilde V(k) \hat c(-{\bf k}),
 \end{equation}
where $c=\zeta_1-\zeta_2$, $r=|{\bf r}|, k=|{\bf k}|$ and by $V$ we denote the product of the interactions and the pair distribution function, $V(r)=\Big(\frac{6}{\pi}\Big)^2u(r)\theta(r-1)$. The factor $\Big(\frac{6}{\pi}\Big)^2$ is present because the volume fraction rather than density is used in (\ref{U}). We use the tilde for functions in Fourier representation. At this stage we do not specify the form of $V$, and only require that $\tilde V(k)$ takes a global  negative minimum for $k=k_0>0$. With such interactions, the concentration wave with the wavenumber $k_0$ leads to the largest decrease of the internal energy compared to the homogeneous state. This decrease of $U$ competes with  the decrease of the  entropy, for which we assume the same form as for the hard-sphere mixture in the local density approximation,  
\begin{equation}
\label{S}
-TS=\int d{\bf r}\Big[k_BT\Big(
\rho_1({\bf r})\ln (\rho_1({\bf r}))+\rho_2({\bf r})\ln (\rho_2({\bf r}))\Big)+ f_{ex}(\zeta({\bf r}))
\Big],
\end{equation}
 where $k_B$ is the Boltzmann constant, $\rho_i({\bf r})=\frac{6}{\pi}\zeta_i({\bf r})$  is the local dimensionless density of the i-th component, and $\zeta=\zeta_1+\zeta_2$. The first two terms come from the entropy of mixing, and the last term is the  contribution to the free energy associated with the packing of the hard cores. In an open system, the last factors determining the structure are the chemical potentials $\mu_i$ of the two components.
 
 In this MF approximation, the grand potential functional takes in terms of  $c=\zeta_1-\zeta_2$ and $\zeta=\zeta_1+\zeta_2$ the following form 
 \begin{equation}
 \label{Om}
 \Omega^{MF}[c,\zeta]= U[c] -TS[c,\zeta]-\int d{\bf r}(\mu_+\zeta({\bf r}) +\mu_-c({\bf r}) ),
 \end{equation}
 where $\mu_+=\frac{3}{\pi}(\mu_1+\mu_2)$ and $\mu_-=\frac{3}{\pi}(\mu_1-\mu_2)$, and $U$ and $S$ are given in (\ref{U}) and (\ref{S}), respectively.
 In equilibrium, $c$ and $\zeta$ correspond to the minimum of $\Omega^{MF}[c,\zeta]$ for given $T,\mu_+,\mu_-$. In this work we neglect the fluctuation contribution to the grand potential that can play a significant role for weakly ordered phases. The role of the variance of the local concentration will be studied in future works.

 \subsection{Boundary of stability of the homogeneous phase in the mean-field approximation}
 \label{mf:bs}
 At high temperature, the system is homogeneous and $c({\bf r})=\bar c$ and $\zeta({\bf r})=\bar \zeta$. When $T$ decreases, 
 oscillatory perturbations of the concentration, $c-\bar c\propto\cos(xk_0)$ and superposition of such waves in different directions can induce instability of the disordered homogeneous 
 phase,  because such waves  lead to the largest decrease of $U$. Although the volume-fraction waves are not directly energetically favored (see (\ref{U})), the coupling between $c$ and $\zeta$ in the entropic contribution to $\Omega$ can support such waves.
 
  Let us consider periodic perturbations about the space-averaged concentration and volume fraction, $\bar c$ and $\bar\zeta$, of the form
 \begin{equation}
 \label{PhPs}
 c({\bf r})=\bar c +\Phi g({\bf r}),\hskip1cm \zeta({\bf r})=\bar\zeta+\Psi g({\bf r}),
 \end{equation}
 where $g({\bf r})$ is a superposition of the plane waves with the wave number $k_0$ in different directions,  $\int_{V_u} d{\bf r} g({\bf r})=0$ and $V_u^{-1}\int_{V_u} d{\bf r} g^2({\bf r})=1$, with $V_u$ denoting the volume or the area  of the unit cell of the periodic structure in three or in two dimensional systems, respectively.
  Here we limit ourselves to 2D patterns expected for particles confined in narrow slits or embedded in bilayers. 
  
In the one-component system with the SALR interactions, stripes and hexagonal arrangement of clusters or voids were found in the previous studies of the  2D models~\cite{imperio:04:0,archer:08:0,almarza:14:0}. 
In the symmetrical binary mixture with $\bar c=0$, alternating stripes of the first and the second component were observed for this type of interactions in Ref.\cite{patsahan:21:0}. For some range of $\bar c\ne 0$, the hexagonal arrangement of clusters of the minority component filling the voids formed by the majority component can be expected, based on the layers adsorbed at a surface attracting particles of the first species, and on the results obtained for a triangular lattice model of a similar mixture~\cite{litniewski:21:0,virgiliis:23:0}.
In principle, a chess-board pattern can  occur in addition to the alternating stripes for $\bar c=0$. 
We assume that such patterns can be stable or metastable in the considered mixture, and shall verify this assumption by MC simulations at different parts of the phase space.
The  $g({\bf r})$ functions for the above patterns are shown in Fig.\ref{fig:patterns}. The mathematical formulas for $g({\bf r})$ describing the 2D patterns 
are given in Appendix. Note that different combinations of signs and magnitudes of $\Phi,\Psi$ can correspond to different hexagonal patters for the same form of $g({\bf r})$, namely to clusters or bubbles in the one-component system, or to clusters of the minority component in the liquid of the majority component. 
\begin{figure}
\includegraphics[scale=0.23]{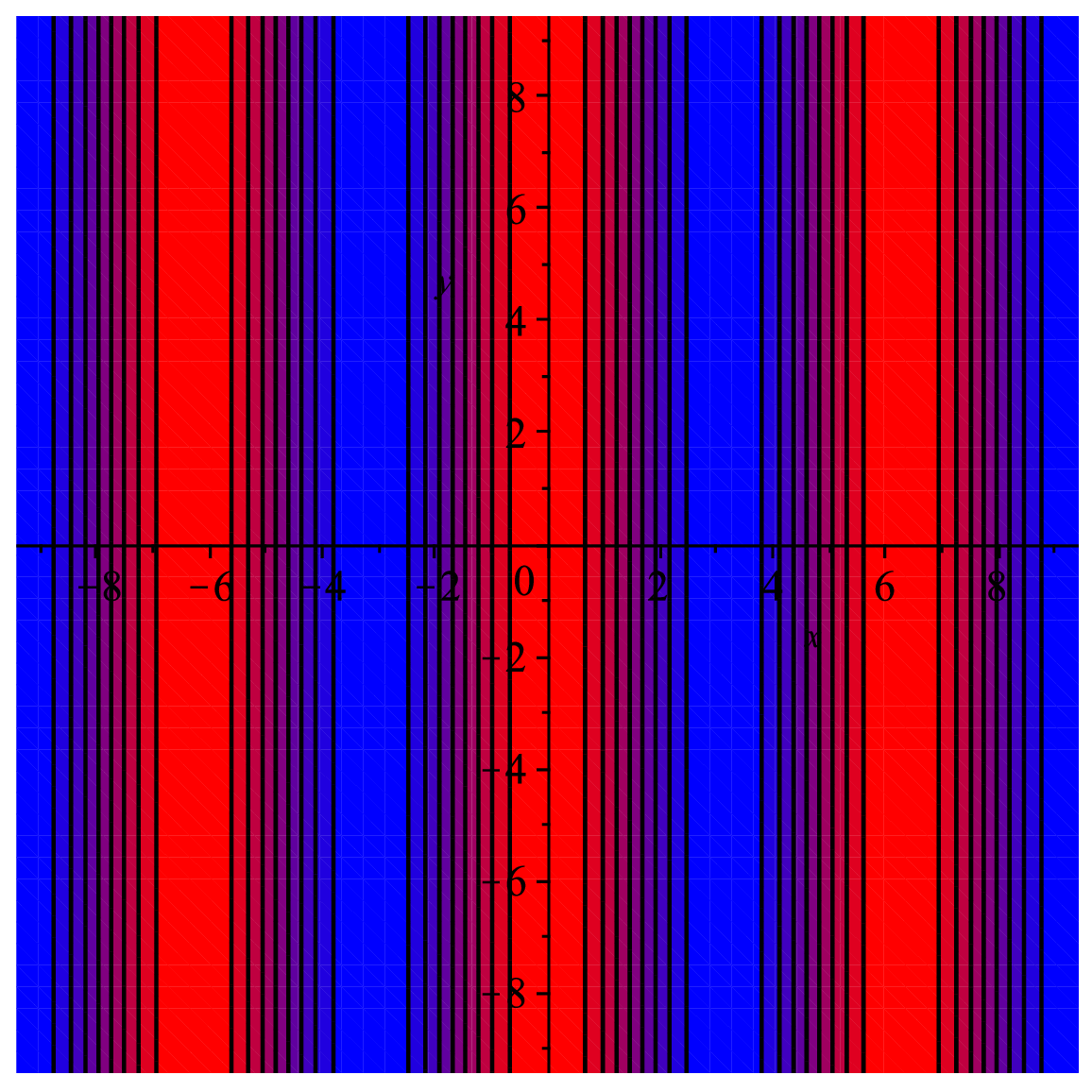}
\includegraphics[scale=0.23]{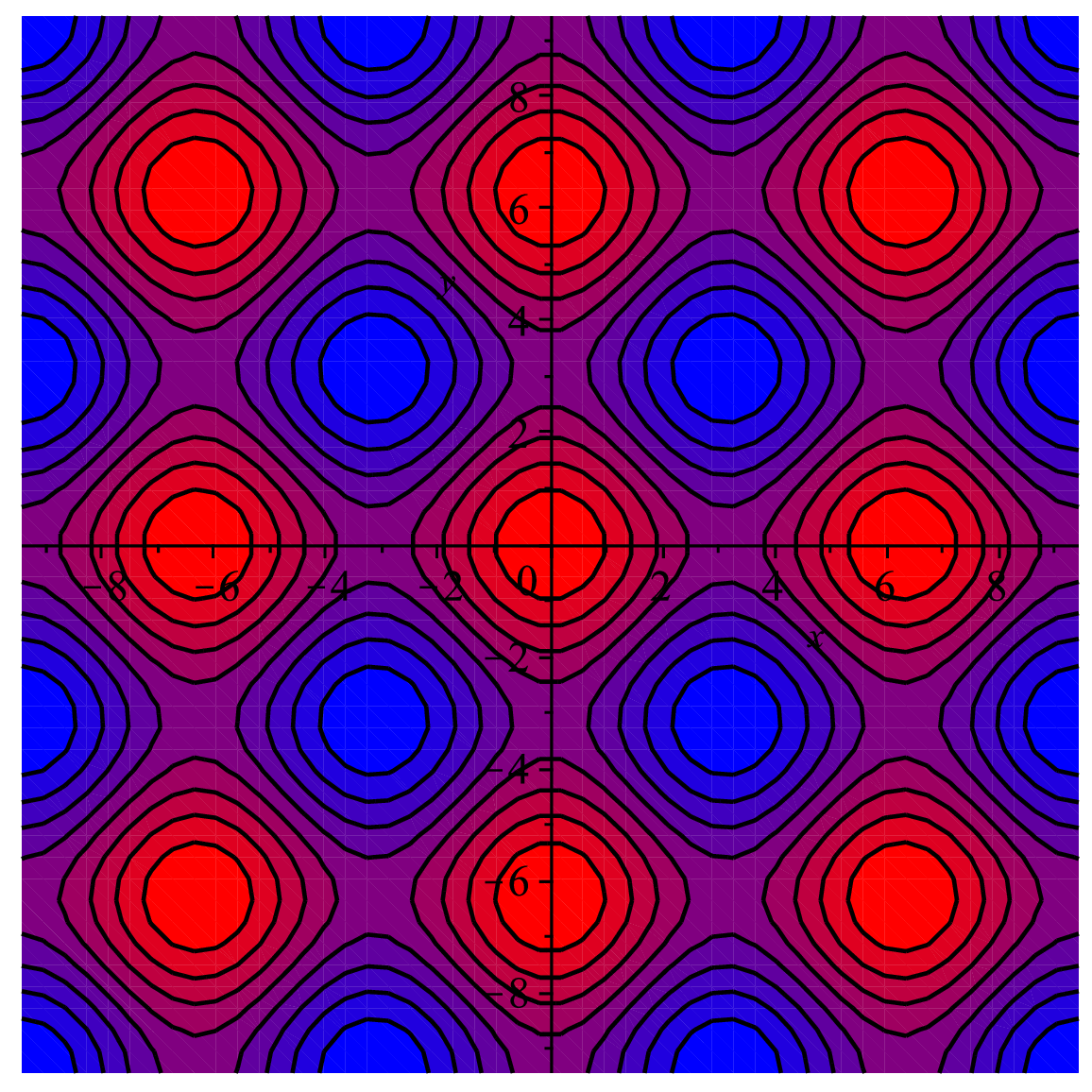}
\includegraphics[scale=0.23]{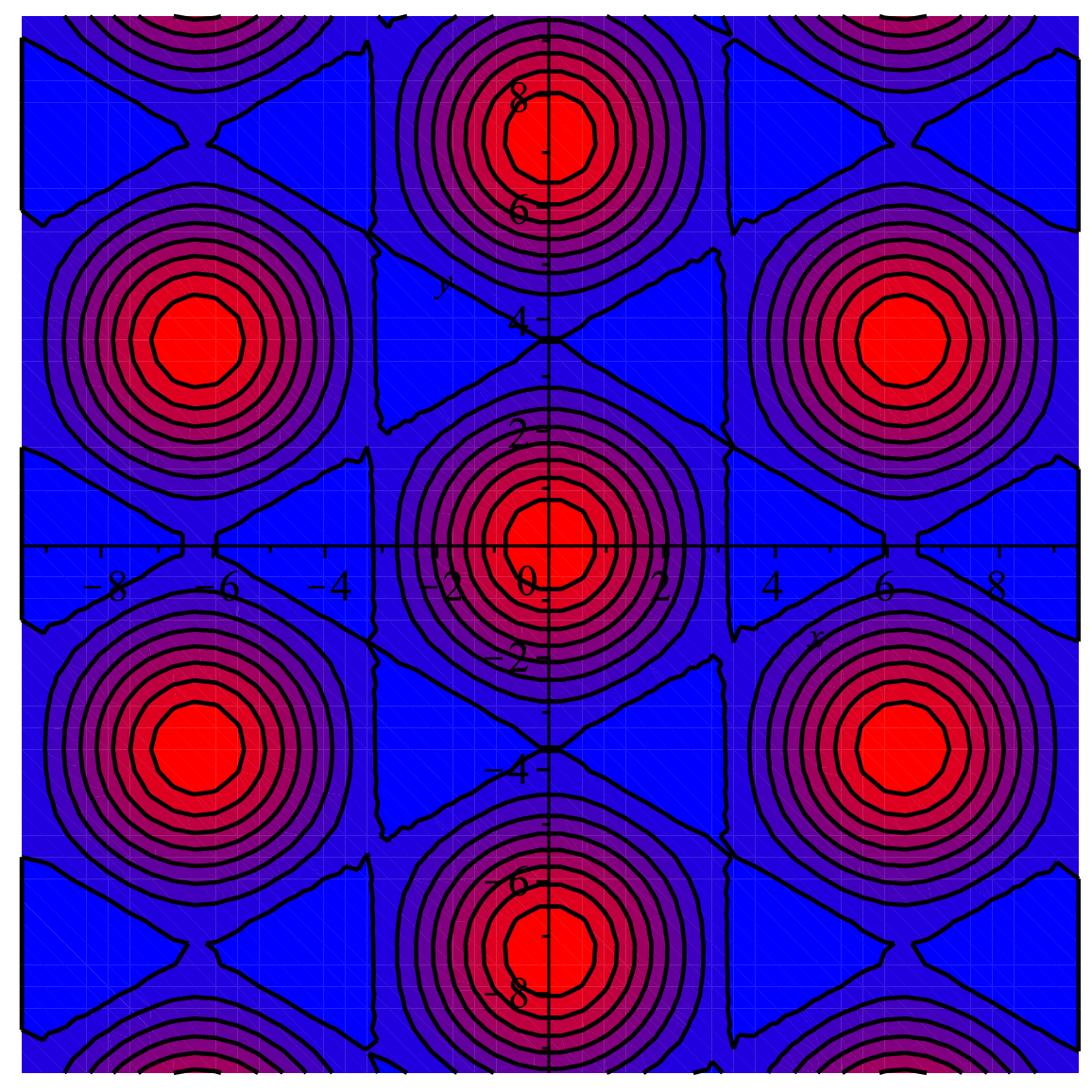}
\vskip0.5cm
\caption{The contour plots of the  considered forms of the $g$ functions. From the left to the right figure, stripes (L phase), chess-board and the hexagonal (H) patterns are shown. Red and blue color corresponds to $g({\bf r})>0$ and $g({\bf r})<0$, respectively. Note that when the amplitude changes sign, only the first two patterns remain the same (up to translation). In the one-component case, the last figure represents a hexagonal arrangement of clusters, and  $-g({\bf r})$ represents  a hexagonal arrangement of bubbles. }
\label{fig:patterns}
\end{figure}

  In the case of 2D patterns, we consider the grand potential per unit area, $\omega=\Omega^{MF}/A$.  For $c$ and $\zeta$ of the form (\ref{PhPs}), the grand potential per unit area takes the form
  \begin{eqnarray}
  \label{bo}
  \beta\omega(\bar c,\bar\zeta,\Phi,\Psi)&=&\frac{3}{\pi}\Bigg[
(\bar\zeta+\bar c )\ln (\bar\zeta+\bar c )+(\bar\zeta-\bar c )\ln  (\bar\zeta-\bar c )+2\ln\left(\frac{3}{\pi}\right) \bar\zeta
  \Bigg]+\beta f_{ex}(\bar\zeta)
  \\
  \nonumber
& +&\frac{\beta}{2}\tilde V(0)\bar c^2 -\beta\mu_+\bar\zeta-\beta\mu_-\bar c  +\Delta \beta\omega(\bar c,\bar \zeta,\Phi,\Psi),
\end{eqnarray}  
where $\Delta \beta\omega$ is associated with the periodic deviations of $c$ and $\zeta$ from the space-averaged values. $\Delta \beta \omega$ can be Taylor-expanded in terms of $\Phi,\Psi$. In order
 to determine the boundary of stability of the homogeneous phase, it is sufficient to truncate  this Taylor expansion
at the second order-terms,
\begin{equation}
\label{bDo}
\Delta\beta\omega=\frac{1}{2}\Bigg( \frac{6\bar\zeta}{\pi(\bar\zeta^2-\bar c^2)}+\beta \tilde V(k_0)\Bigg)\Phi^2+
\frac{1}{2}\Bigg( \frac{6\bar\zeta}{\pi(\bar\zeta^2-\bar c^2)}+A_2(\bar\zeta)\Bigg)\Psi^2-\frac{6\bar c}{\pi(\bar\zeta^2-\bar c^2)}\Phi\Psi+h.o.t.,
\end{equation}
where h.o.t. means higher-order terms, and 
\begin{equation}
\label{An}
A_n(\zeta)=\frac{d^n \beta f_{ex}(\zeta)}{d \zeta^n}.
\end{equation}
In our computations,  we assume the Carnahan-Starling form for $\beta f_{ex}$~\cite{Carnahan1969}.
 At the boundary of stability of the homogeneous phase, the determinant of the matrix of the second derivatives of $\beta\Delta\omega$ with respect to $\Phi$ and $\Psi$ vanishes, and the corresponding $\lambda$-surface is given by
 \begin{equation}
 T^*_{\lambda}(\bar c,\bar\zeta)=\frac{\pi}{6}\frac{\bar\zeta+\frac{\pi}{6}(\bar\zeta^2-\bar c^2)A_2(\bar\zeta)}{1+\frac{\pi}{6}\bar\zeta A_2(\bar\zeta)},
 \end{equation}
where  $T^*=k_BT/|\tilde V(k_0)|$ is the dimensionless temperature. At the state points $(\bar c,\bar\zeta,T^*)$ with $T^*<T_{\lambda}^*(\bar c,\bar\zeta)$, the homogeneous phase is unstable, and  either one of the periodic  structures is present, or different phases coexist. 
 Note that  when the dimensionless temperature is given by the ratio of the thermal energy and the energy gain due to the excitation of the concentration wave with the optimal wavelength and unitary amplitude, the $\lambda$-surface is universal, i.e. independent of the form of the interactions. This universality resembles the law of corresponding states in the van der Waals theory.

\begin{figure}
\includegraphics[scale=0.55]{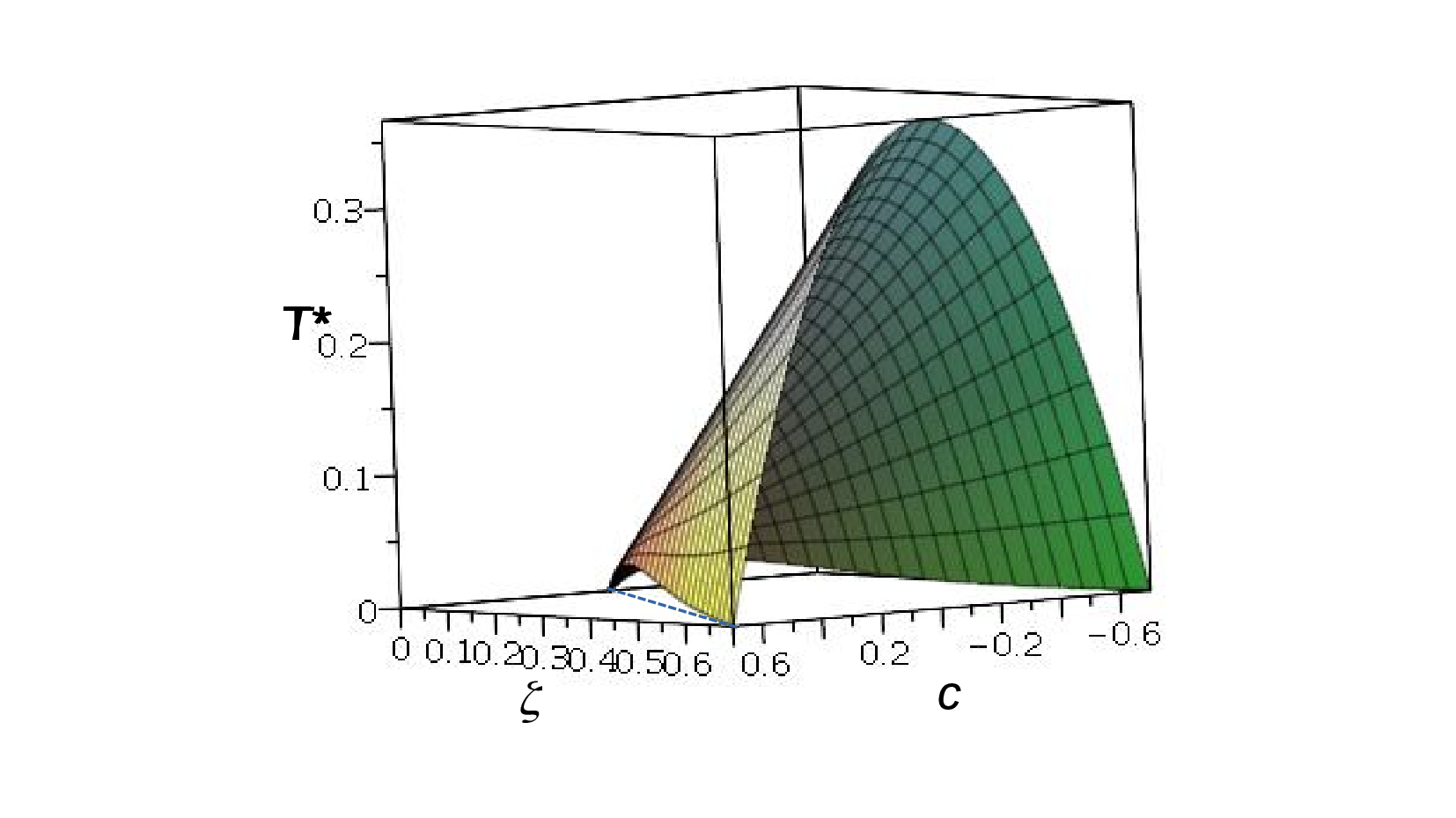}
\includegraphics[scale=0.38]{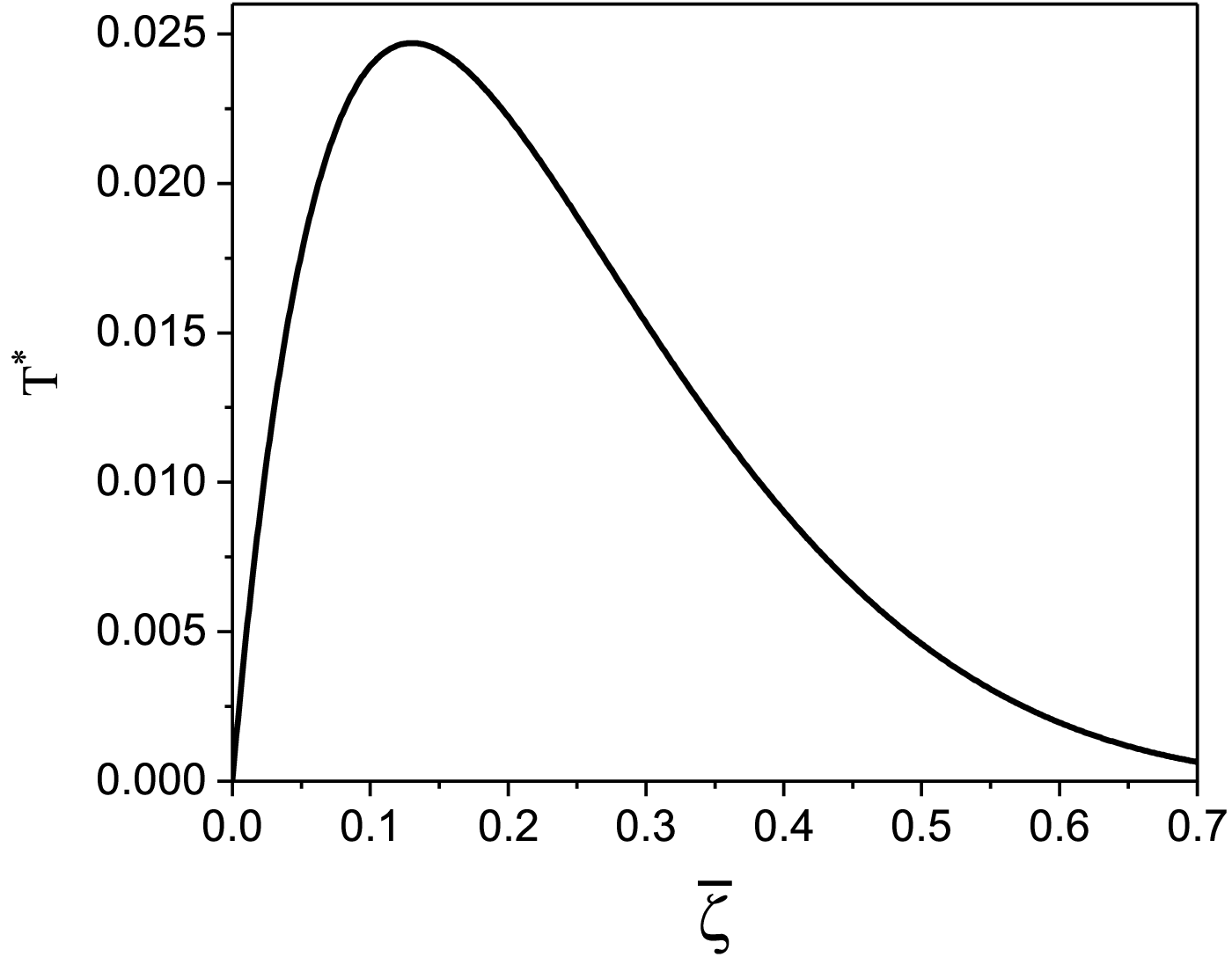}
\caption{The $\lambda$-surface in the two-component system with competing interactions (top), and the $\lambda$-line in the absence of the second component (bottom). The blue dashed line in the top figure indicates the one-component system with $\bar c=\bar\zeta$. }
\label{fig:spinline}
\end{figure}

It is instructive to compare $T_{\lambda}^*$ for the one-component system with $\bar\zeta=\bar c$ (i.e. with $\zeta_2=0$) where $T_{\lambda}^*=\pi\bar\zeta/(6+\pi A_2(\bar\zeta)\bar\zeta)$, and the symmetrical mixture with $\bar c=0$ where $T_{\lambda}^*=\pi\bar\zeta/6$ (see Fig.\ref{fig:spinline}).  Note that the ordered phases can be present at much higher temperature when the second component is added.  For given $\bar\zeta$, the temperature at  the instability of the homogeneous phase takes the maximum when both components are in equal proportions. Moreover, in the one component case the temperature at the instability takes the maximum
 at much smaller $\bar \zeta$ than in the mixture with $\bar c=0$, where it linearly increases with $\bar \zeta$.
 
\begin{figure}
\includegraphics[scale=0.22]{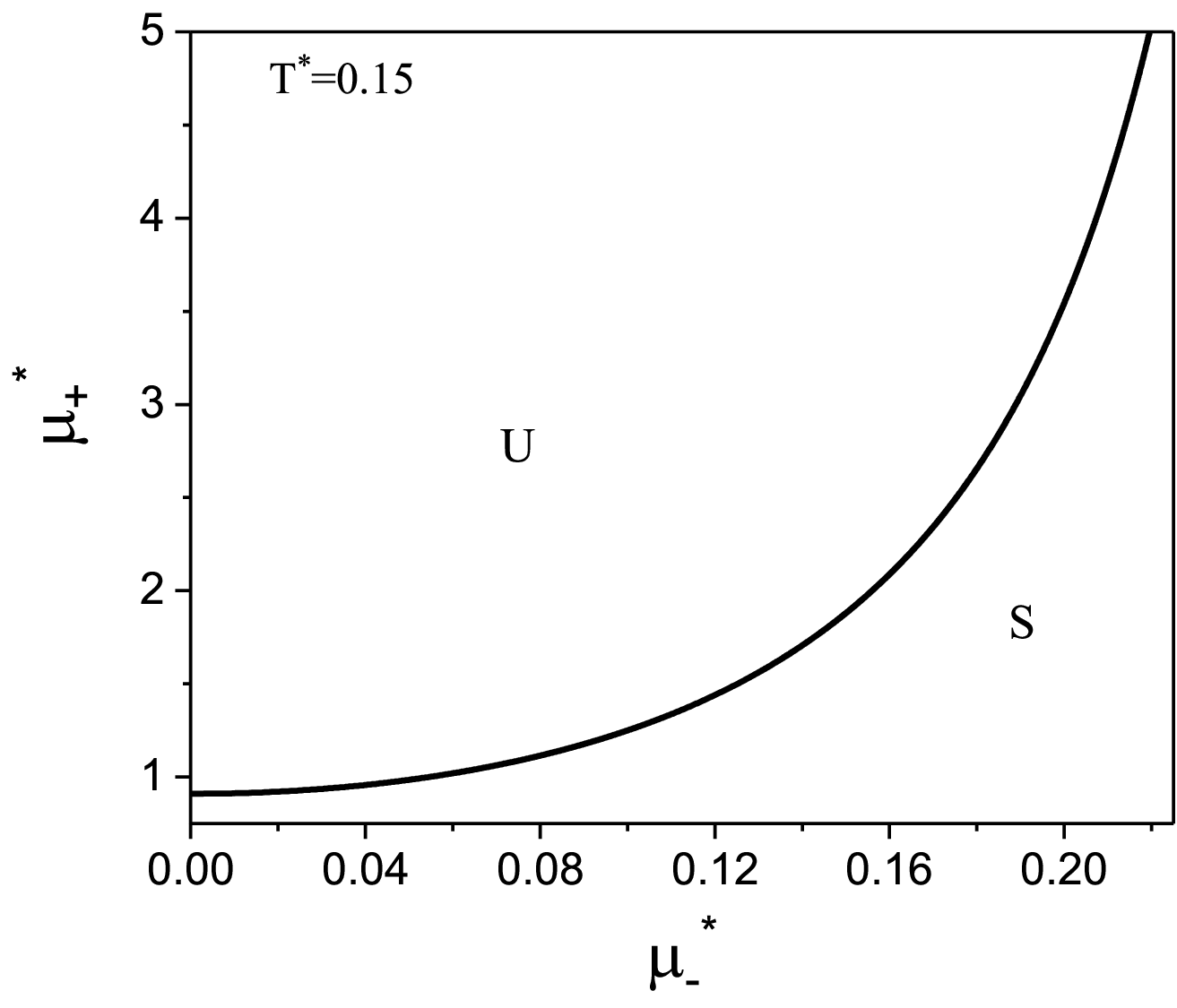}
\includegraphics[scale=0.22]{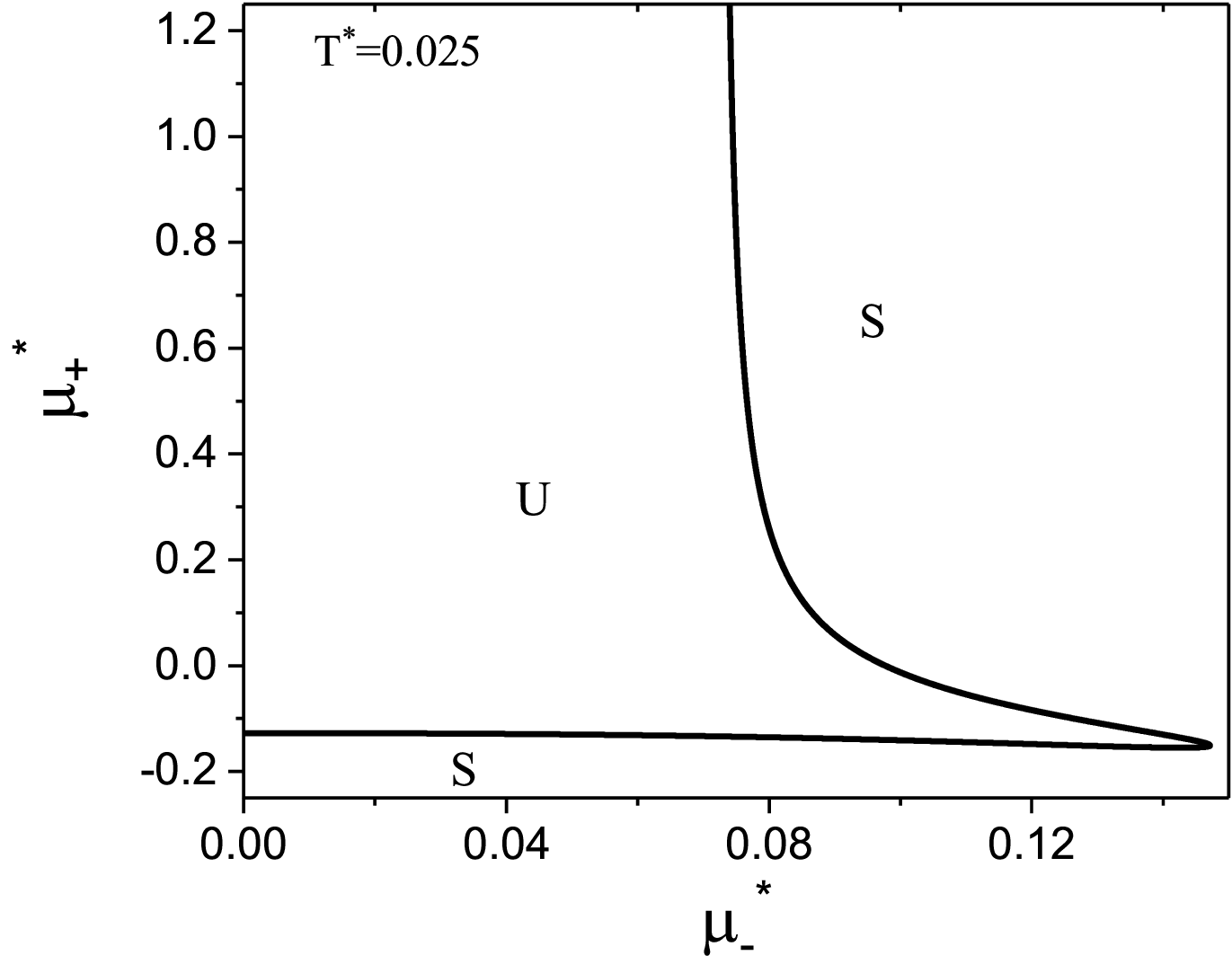}
\includegraphics[scale=0.22]{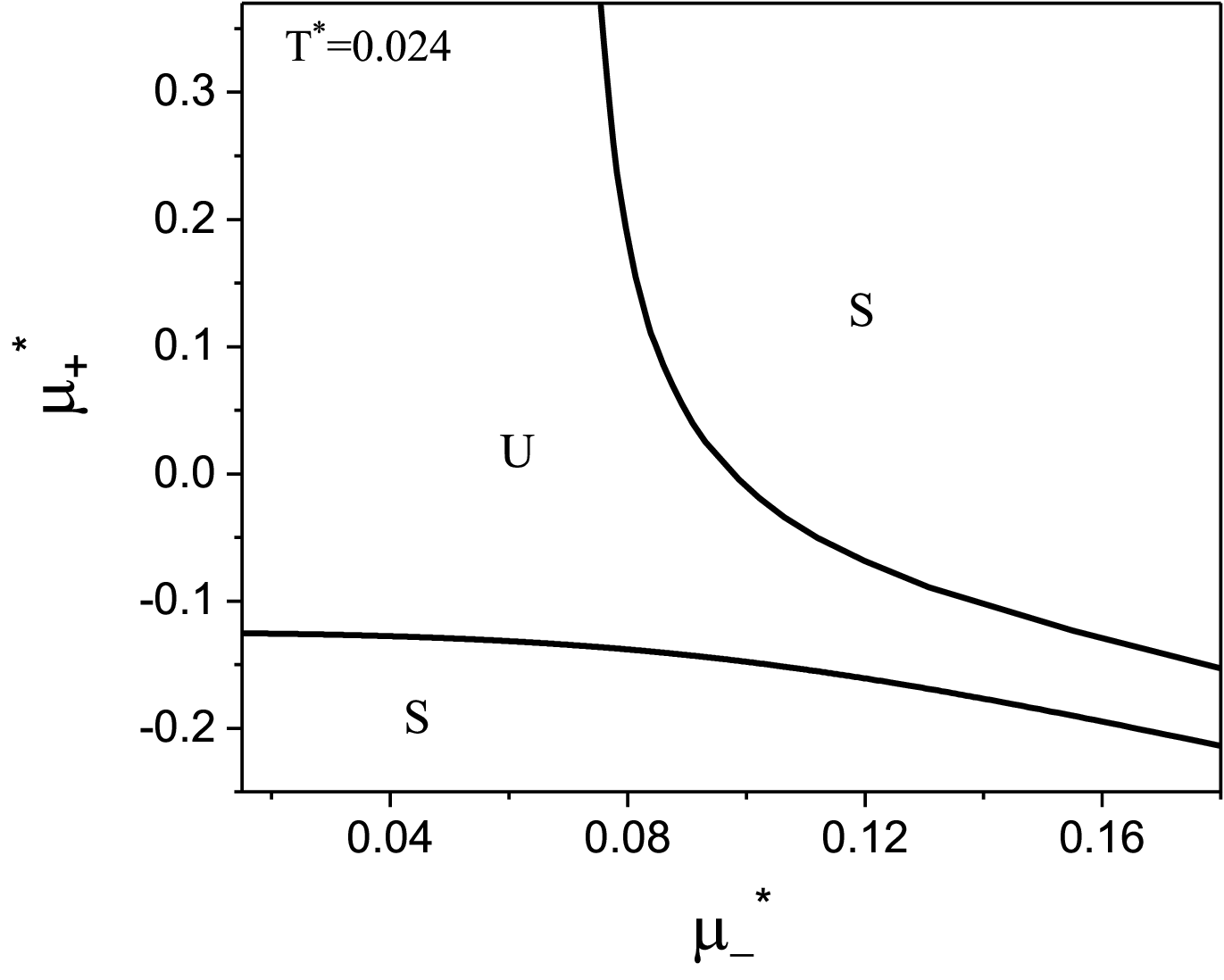}
\vskip0.5cm
\caption{The cross-sections through the $\lambda$-surface for fixed temperature. Only $\mu_{-}^{*}> 0$ is shown, since the vertical axis is the  symmetry axis of the full diagram. From the left to the right figure, $T^*=0.15,0.025, 0.024$. U and S refer to unstable and stable disordered phase, respectively. 
$T^*=k_BT/|\tilde V(k_0)|$, and  $\mu_{\pm}^*= \mu_{\pm}/|\tilde V(k_0)|$, where $\mu_{\pm}=\frac{3}{\pi}(\mu_1\pm\mu_2)$, and $\mu_i$ is the chemical potential of the i-th species. }
\label{fig:spinline_mu}
\end{figure}

 It is also interesting to analyze the $\lambda$-surface in the $(\mu_-,\mu_+,T^*)$ variables, since in these variables we construct the phase diagram.  From the minimum condition of the grand potential with respect to $c$ and $\zeta$ in the homogeneous phase, including its boundary of stability,  we have the relations
 \begin{eqnarray}
\label{bmu+D}
\beta\mu_+^D=\frac{3}{\pi}\ln(\bar\zeta+\bar c)+A_1(\bar\zeta)+\frac{6}{\pi}\Bigg[\ln\Big(\frac{3}{\pi}\Big)+1\Bigg] +\frac{3}{\pi}\ln(\bar\zeta-\bar c)
\end{eqnarray}
and
 \begin{eqnarray}
\label{bmu-D}
\beta\mu_-^D=\frac{3}{\pi}\ln(\zeta+\bar c)+\beta\tilde V(0)\bar c-\frac{3}{\pi}\ln(\bar\zeta-\bar c),
\end{eqnarray}
where the superscript D denotes the disordered phase with $\Phi=\Psi=0$. 
The  cross-sections of the $\lambda$-surface  for fixed $T^*=0.15,0.025,0.024$ are shown in Fig.\ref{fig:spinline_mu}. We can see that at high $T^*$, there is a single disordered phase, i.e. there exists a continuous path from the dilute gas of mixed components to the dense one-component liquid. At low $T^*$, however, the dilute two-component gas is separated from the one-component disordered liquid by the one-component ordered phases.

 \subsection{One-shell approximation for the periodic phases}
 \label{sec:one_shell}
In the  one-component SALR model, the density in the ordered phases  at  relatively high $T^*$
can be approximated by a superposition of plane waves, but for low temperatures the shape of $\zeta({\bf r})-\bar\zeta$ strongly deviates from the sinusoidal function~\cite{pini:17:0}.
Similar behavior can be expected in the inhomogeneous mixture.  The purpose of this work, however, is determination of the $(\mu_-^*,\mu_+^*,T^*)$ and $(\bar c,\bar\zeta,T^*)$ phase diagrams on a qualitative level, and in this first study of the global phase diagram, we restrict ourselves to the one-shell approximation~(\ref{PhPs}). Instead of $\beta\mu_{\pm}$ convenient in the theoretical formulas, for presentation of the results we introduced the dimensionless chemical potentials independent of $T$ by $\mu_{\pm}^*= \mu_{\pm}/|\tilde V(k_0)|$, where $\mu_{\pm}=\frac{3}{\pi}(\mu_1\pm\mu_2)$ and $\mu_i$ denotes
the chemical potential of the i-th species.

In order to get the qualitative phase diagram, we  truncate the Taylor expansion of $\beta\Delta\omega$ in terms of $\Phi$ and $\Psi$ at the fourth-order terms, by which the stability of the functional (\ref{bo}) with (\ref{bDo})  for  $T^*<T^*_{\lambda}$ is restored. The truncation of the Taylor expansion is valid for small $\Phi, \Psi$, and is consistent with the one-shell approximation. The explicit expressions  in the $\varphi^4$ theory for the higher-order terms in $\Delta\beta\omega$ (see (\ref{bDo}))  can be obtained easily, and are not given here.

In the stable or metastable phase,
 the derivative of $\beta\omega(\bar c,\bar\zeta,\Phi,\Psi)$ with fixed $T^*,\mu_+^*,\mu_-^*$ vanishes. We need to solve 4 equations, $\partial \beta\omega/\partial \bar c =\partial \beta\omega/\partial \bar \zeta=\partial \beta\omega/\partial \Phi= \partial \beta\omega/\partial \Psi= 0$ with fixed $T^*,\mu_+^*,\mu_-^*$ to determine $\bar c,\bar\zeta,\Phi,\Psi$ in the stable or metastable phase for the given thermodynamic state. Two phases coexist for the state-point $T^*,\mu_+^*,\mu_-^*$, when $\beta\omega$ in these phases takes the same value. At the minimum, the grand potential per unit area takes in our approximation the form
\begin{eqnarray}
\label{boXZ}
\beta\bar\omega=-\frac{\beta}{2}\tilde V(0) \bar c^2 +\beta f_{ex}(\bar \zeta) -\Bigg(\frac{6}{\pi}+A_1
\Bigg)\bar \zeta +\frac{\beta \tilde V(k_0)}{2} \Phi^2
\\
\nonumber + \frac{(A_2-\bar\zeta A_3)}{2}\Psi^2 +\frac{(A_3-\bar\zeta A_4)\kappa_3}{3!}\Psi^3+\frac{(A_4-\bar\zeta A_5)\kappa_4}{4!}\Psi^4
\\
\nonumber +
\frac{3}{\pi}\Big(X^2+Z^2
\Big)-\frac{3\kappa_3}{2\pi}\Big(X^3+Z^3
\Big)+\frac{\kappa_4}{\pi}\Big(X^4+Z^4
\Big)
\end{eqnarray}
where $\kappa_n=V_u^{-1}\int_{V_u} d{\bf r} g^n({\bf r})$  
are the geometric factors characterizing different structures, $A_n$ are defined in (\ref{An}), we introduced the notation
\begin{equation}
 \label{XZ}
 X=\frac{\Phi+\Psi}{\bar\zeta+\bar c},\hskip1cm Z=\frac{\Psi-\Phi}{\bar\zeta-\bar c},
 \end{equation}
 and $\bar c,\bar\zeta,\Phi,\Psi$ satisfy the  equations  $\partial \beta\omega/\partial \bar c =\partial \beta\omega/\partial \bar \zeta=\partial \beta\omega/\partial \Phi= \partial \beta\omega/\partial \Psi= 0$ for the considered state-point $T^*,\mu_+^*,\mu_-^*$. The
 explicit forms of the above equations in our approximate $\varphi^4$ theory are given in Appendix.

  Our general framework with $c\le \zeta$ includes the symmetrical case with $\bar c=0$ already studied in Ref.\cite{patsahan:21:0}, as well as the one component case with $\zeta_2=0$, i.e. $\zeta=c$, studied in Ref.\cite{ciach:10:1}. 
Determination of  $\bar c,\bar\zeta,\Phi,\Psi$ in the one-shell approximation for given $T^*,\mu_+^*,\mu_-^*$ and $\kappa_n$ is in principle an easy algebraic problem. However, as can be seen from (\ref{bo}) and (\ref{bDo}),  when $\bar c\to \bar\zeta$, singularities in $\Delta\beta\omega$ appear. In the one-component system, we have $\Phi=\Psi$ in addition to $\bar c= \bar\zeta$, and the singularities present in individual terms cancel out, giving the grand potential of the form
\begin{equation}
\label{bo1}
\beta\bar\omega_1=\frac{1}{2}\beta\tilde V(0)\bar\zeta^2 +\frac{6}{\pi}\Bigg(\bar\zeta\ln(2\bar\zeta)+\ln\Big(\frac{3}{\pi}\bar\zeta
\Big) \Bigg)+\beta f_{ex}(\bar\zeta)-\beta\bar\mu_1\bar\zeta+\Delta\beta\omega_1,
\end{equation}
where $\bar\mu_1=\mu_++\mu_-$ and 
\begin{equation}
\label{Dbo1}
\Delta\beta\bar\omega_1\approx \frac{1}{2}\Bigg(\frac{6}{\pi\bar\zeta}+A_2(\bar\zeta)-\frac{1}{T^*}
\Bigg)\Phi^2 -\frac{\kappa_3}{3!}\Bigg(\frac{6}{\pi\bar\zeta^2}-A_3(\bar\zeta)
\Bigg)\Phi^3+\frac{\kappa_4}{4!}\Bigg(\frac{12}{\pi\bar\zeta^3}+A_4(\bar\zeta)\Bigg)\Phi^4.
\end{equation}
 The expressions for  $\Phi$ and $\bar \zeta$  with fixed $T^*$ and $\beta\bar\mu_1$ are given in Appendix.
 
 Note that when a small amount of the second component is added,  (i.e.  $\bar c\to\bar\zeta$), the coefficients of the individual terms in $\beta\omega$ are very large (see (\ref{bDo})).
 We  introduced $X$ and $Z$ (see (\ref{XZ})), and transformed the original Taylor expansions to the form where the dependence on $1/(\bar\zeta-\bar c)^n$ is  through $Z^n=[(\Psi-\Phi)/(\bar\zeta-\bar c)]^n$.  
 Since we expect $\Phi\to\Psi$ for  $\bar c\to\bar\zeta$,
  our expression (\ref{boXZ}) and formulas in Appendix  are suitable for numerical calculations.
  
\section{The interactions}
\label{sec:model}
We choose for the interaction $u(r)$ between particles of the same species  the hard-core repulsion plus the double Yukawa potential,
\begin{equation}
\label{Yukawa}
u(r)= -K_1\frac{\exp(-\kappa_1 r)}{r}+K_2\frac{\exp(-\kappa_2 r)}{r},
\end{equation}
with $K_1=1$, $K_2=0.2$, $\kappa_1=1$ and $\kappa_2=0.5$. The interaction between particles of different species is $u_{12}(r)=-u(r)$ for $r>1$. 
The potential (\ref{Yukawa})  will be used in simulations verifying and illustrating our results.

\begin{figure}
\includegraphics[scale=0.3]{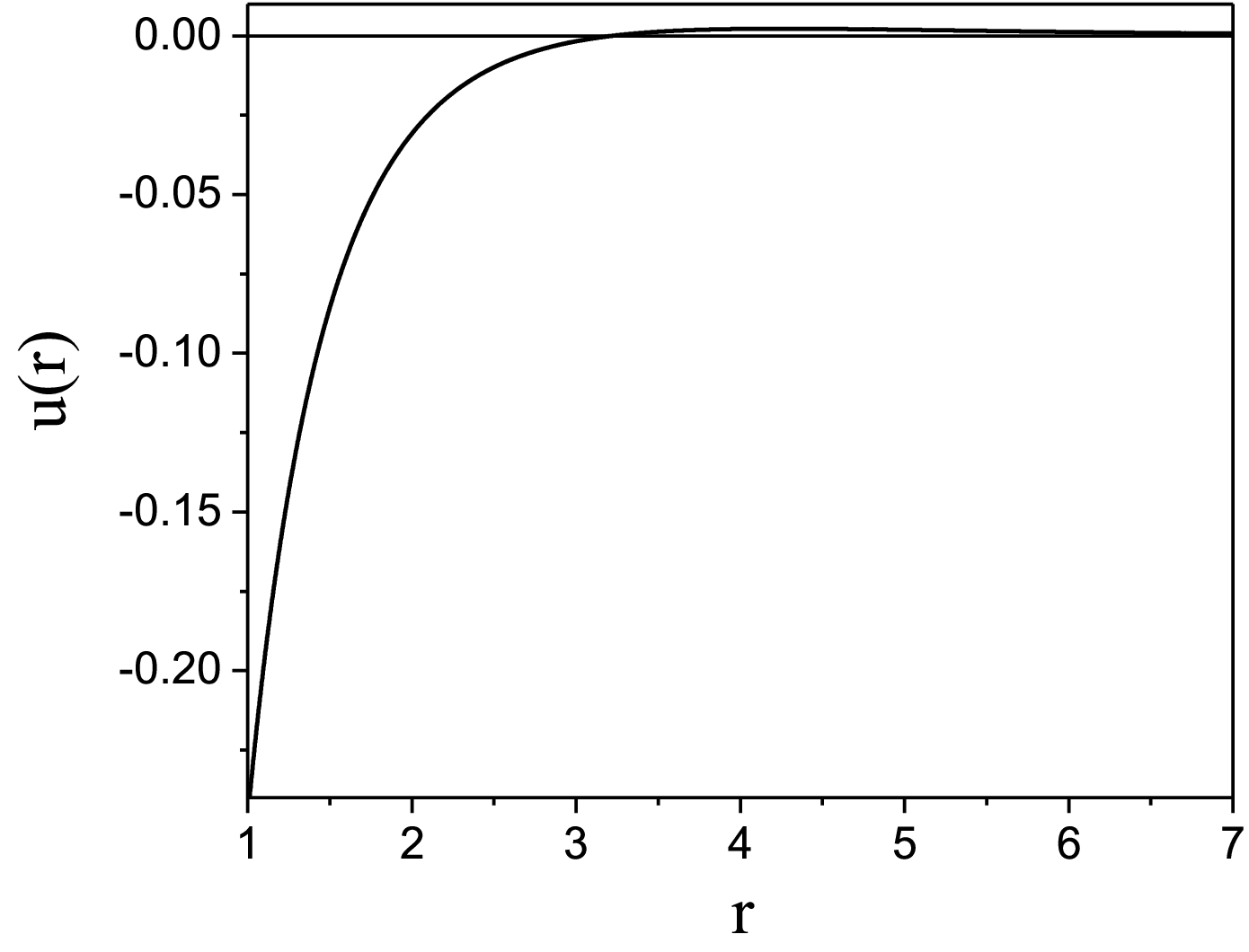}
\includegraphics[scale=0.3]{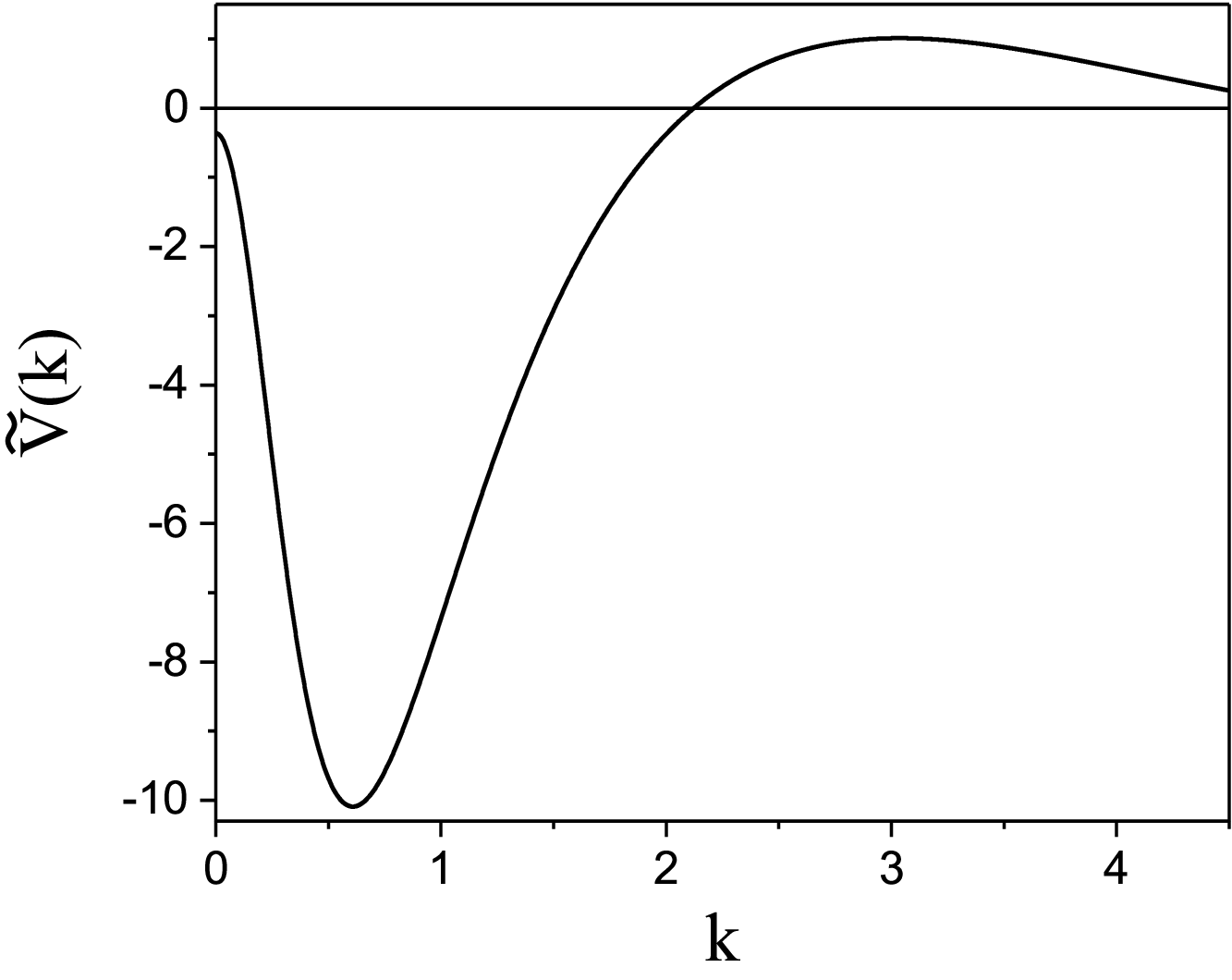}
\caption{Left: the interaction potential $u(r)$ in real space  in units of the attraction strength (see (\ref{Yukawa})). Note the deep minimum at short distances and low but broad maximum for large distances. Right: the Fourier transform  $\tilde V(k)$ of $V(r)=(\frac{6}{\pi})^2u(r)\theta(r-1)$  (see Eq.(\ref{U})). The length unit is the particle diameter $a$, and $k$ is in $1/a$ units.
In simulations, the particles are confined to a 3D  slit with the thickness $1.5 a$,  therefore the Fourier transform is calculated for a 3D system. }
\label{fig:V}
\end{figure}

 As discussed in sec.\ref{mt} and sec.\ref{mf:bs} (see   (\ref{U}), (\ref{bo}) and (\ref{bDo})), 
the phase diagram in our one-shell approximation depends only on 3 parameters characterizing the interaction potential in Fourier representation, $k_0, \tilde V(k_0)$ and $\tilde V(0)$ (recall that $V(r)=(\frac{6}{\pi})^2u(r)\theta(r-1)$). $\pi/k_0$ characterizes the size of self-assembled aggregates, and $|\tilde V(k_0)|$ sets the temperature  scale. The last parameter,  $\tilde V(0)=\int d{\bf r}V(r)$, determines whether in the absence of periodic patterns, the homogeneous mixed phase (for $\tilde V(0)>0$) or separated homogeneous components (for $\tilde V(0)<0$)  are energetically favored.
For the interaction (\ref{Yukawa}),
the relevant parameters take  the values: $k_0\simeq 0.6088,\tilde V(k_0)\simeq -10.092, \tilde V(0)\simeq -0.363$. With the chosen interactions, the size of the clusters is $\pi/k_0\approx 5$, and separation of the homogeneous components is energetically favored over the homogeneous  mixture.
The function $u(r)$ in real space and the function $\tilde V(k)$ in Fourier representation are shown in Fig.\ref{fig:V}. 

\section{The results}
\label{sec:reults}
In this section, we first present the 
 phase diagram in the one-component system. Next, we show the estimated $(\bar c,T^*)$ diagram for fixed large value of $\bar\zeta$ and relatively large $T^*$, based only on the minimization of  $\Delta\beta\omega$ with respect to $\Phi,\Psi$. 
Finally, we present the $(\mu_-^*,\mu_+^*)$ diagrams for a few selected temperatures, 
 obtained from the minimization of $\beta\omega$ for fixed $\mu_-^*,\mu_+^*$ and $T^*$.
 The corresponding $(\bar c, \bar \zeta)$ diagrams are  shown too. We choose $T^*$ above and below the boundary of stability of the homogeneous phase in the one-component system. Our results are verified by MC simulations, and representative configurations are shown for selected state points.
 
\subsection{Theoretical results}
\label{sec:theory_results}
\begin{figure}
\includegraphics[scale=0.3]{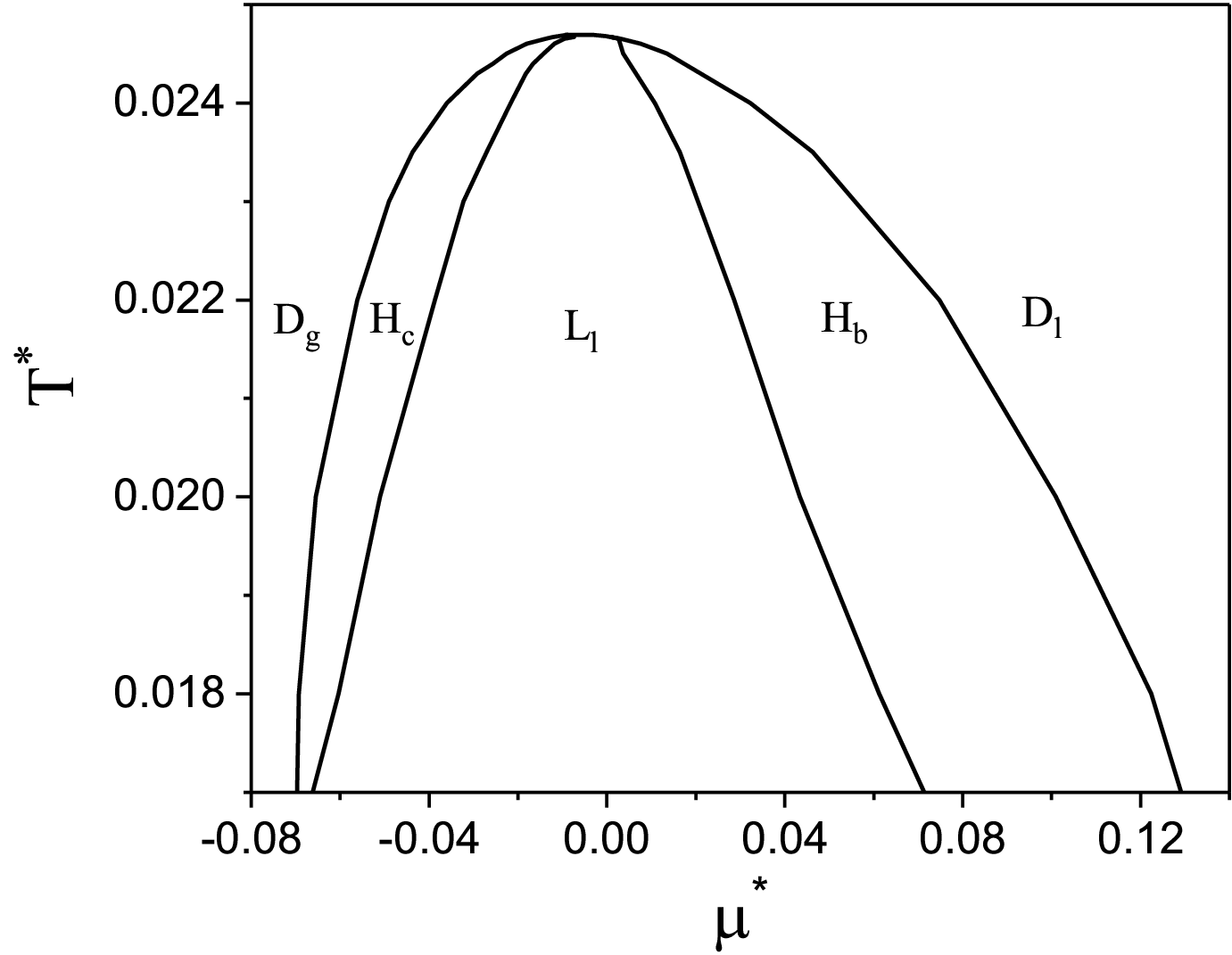}
\includegraphics[scale=0.3]{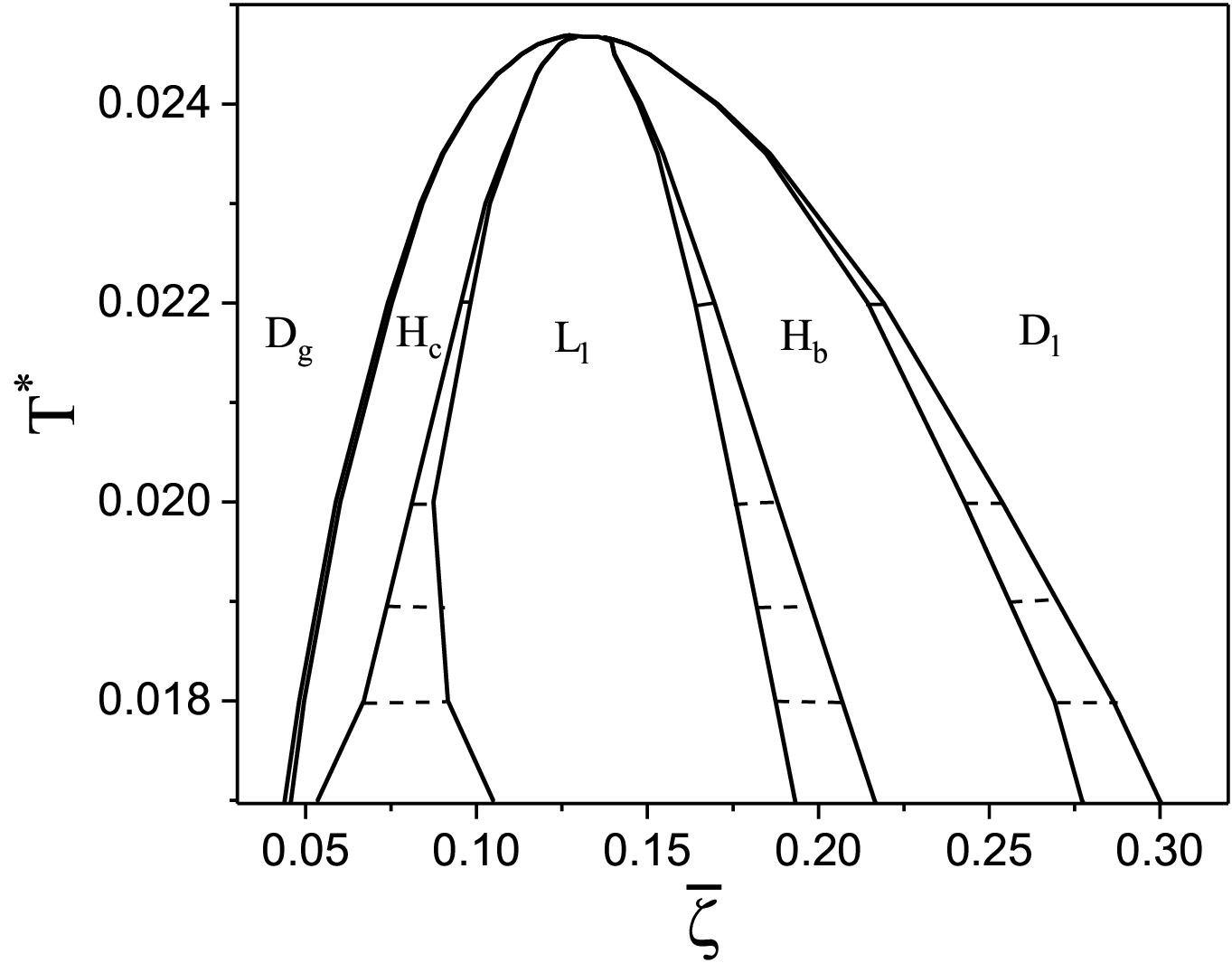}
\caption{The phase diagram in the one-component SALR system obtained by minimization of the grand potential (Eq.~(\ref{bo1})). Left and right panels show the diagram in the $(\mu^*,T^*)$ and $(\bar\zeta,T^*)$ variables. 
${\rm D_g,H_c,L_1,H_b}$ and ${\rm D_l}$ denote the disordered gas, hexagonal pattern of clusters, stripes separated by empty layers, hexagonal pattern of bubbles and disordered liquid, respectively. $T^*=k_BT/|\tilde V(k_0)|$ and $\mu^*= \mu_{+}^{*}+\mu_{-}^*$.}
\label{fig:1com_diag}
\end{figure}

The sequence of ordered phases in the one-component system was previously determined within our mesocopic theory  by considering only the excess grand potential associated with periodic oscillations of the volume fraction, $\Delta\beta\omega_1$ (see (\ref{Dbo1})).  Here we  take into account different densities in the coexisting phases. The obtained phase diagram is shown in Fig.\ref{fig:1com_diag}. Note that the two-phase regions are quite narrow. 

\begin{figure}
\includegraphics[scale=0.3]{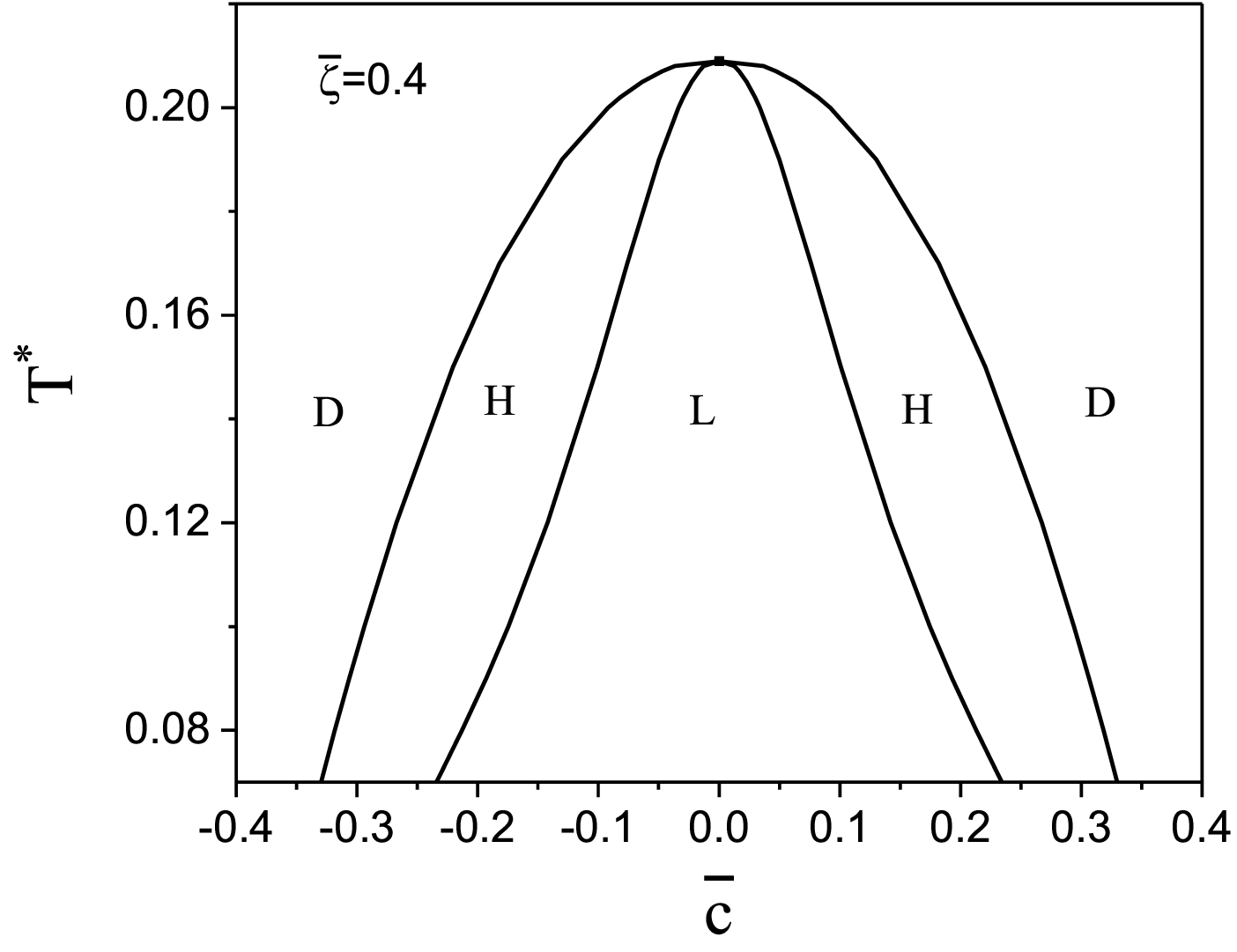}
\caption{The $(\bar c,T^*)$  phase diagram  obtained by minimization of $\Delta\beta\omega$ (see Eq.(\ref{bo})) with respect to $\Phi,\Psi$  in our $\varphi^4$ approximation for $\bar\zeta=0.4$. D, H, L denote the disordered phase rich in the majority component, the hexagonal phase of clusters of the minority component in the liquid of the majority component, and the lamellar phase of alternating stripes of the two components, respectively.
}
\label{fig:c-T_diag}
\end{figure}

\begin{figure}
\includegraphics[scale=0.3]{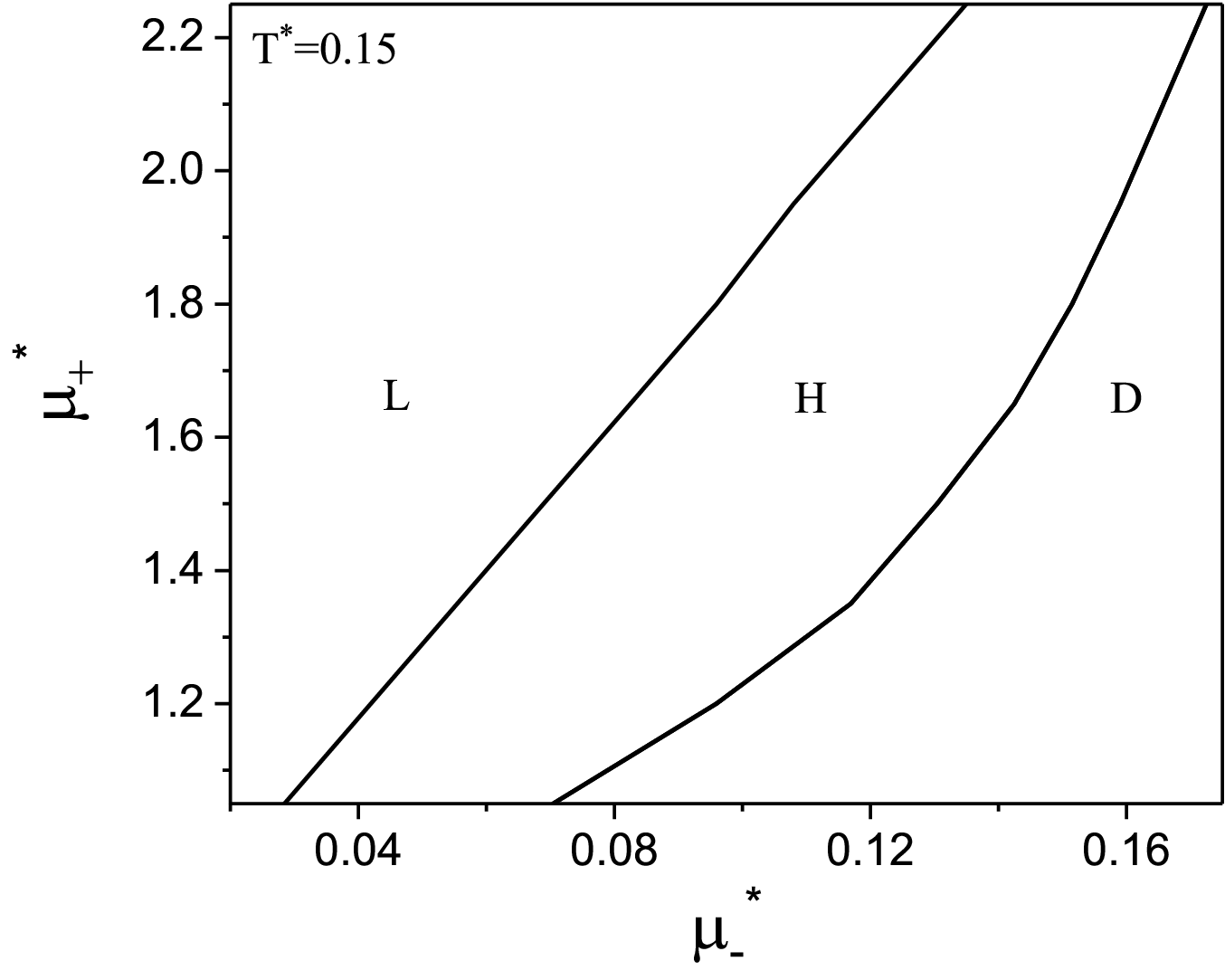}
\includegraphics[scale=0.3]{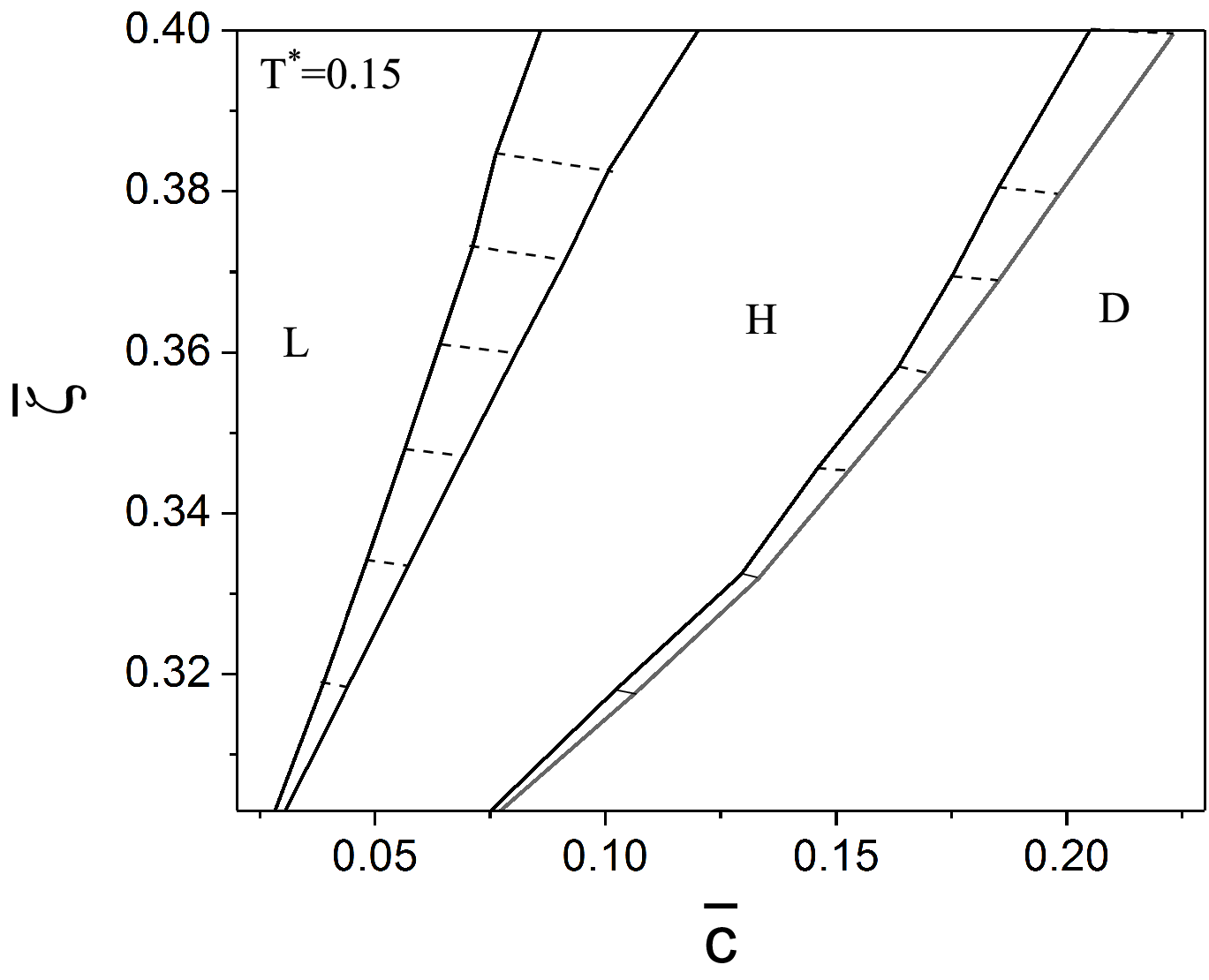}
\caption{The cross-sections through the phase diagram for fixed temperature $T^*=0.15$. Only $\mu_{-}^{*}>0$ and $\bar c>0$ are shown, since the vertical axes are the  symmetry axes of the full diagrams. Left and right figure shows the diagram in the $(\mu_-^*,\mu_+^*)$ and $(\bar c,\bar\zeta)$ variables, respectively. Dashed lines in the right figure are the tie lines. L, H, D denote the lamellar (stripe), hexagonal and disordered phases, respectively. $T^*=k_BT/|\tilde V(k_0)|$ and  $\mu_{\pm}^*= \mu_{\pm}/|\tilde V(k_0)|$, where $\mu_{\pm}=\frac{3}{\pi}(\mu_1\pm\mu_2)$, and $\mu_i$ is the chemical potential of the i-th species. 
}
\label{fig:diag_T0_15}
\end{figure}

\begin{figure}
\includegraphics[scale=0.32]{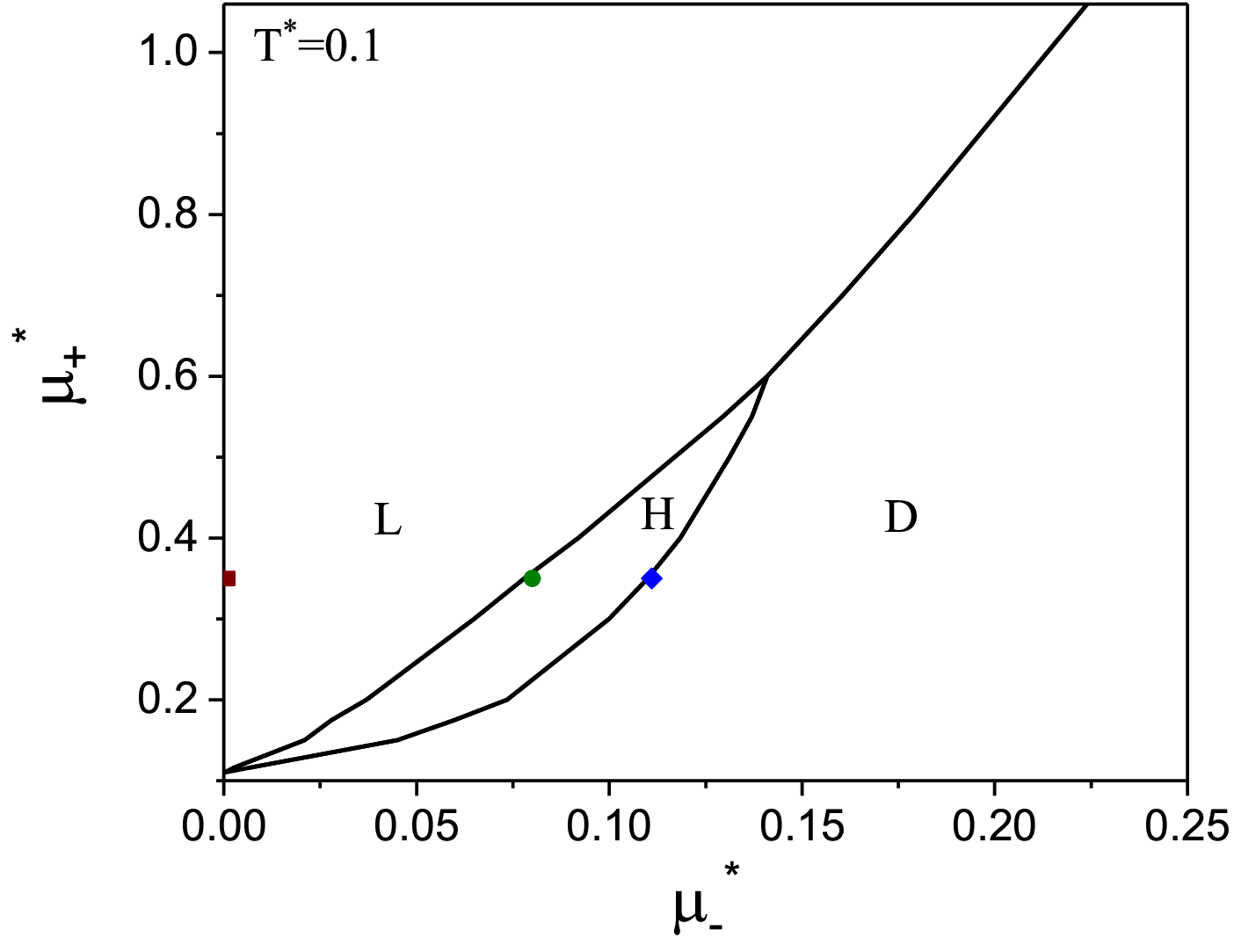}
\includegraphics[scale=0.32]{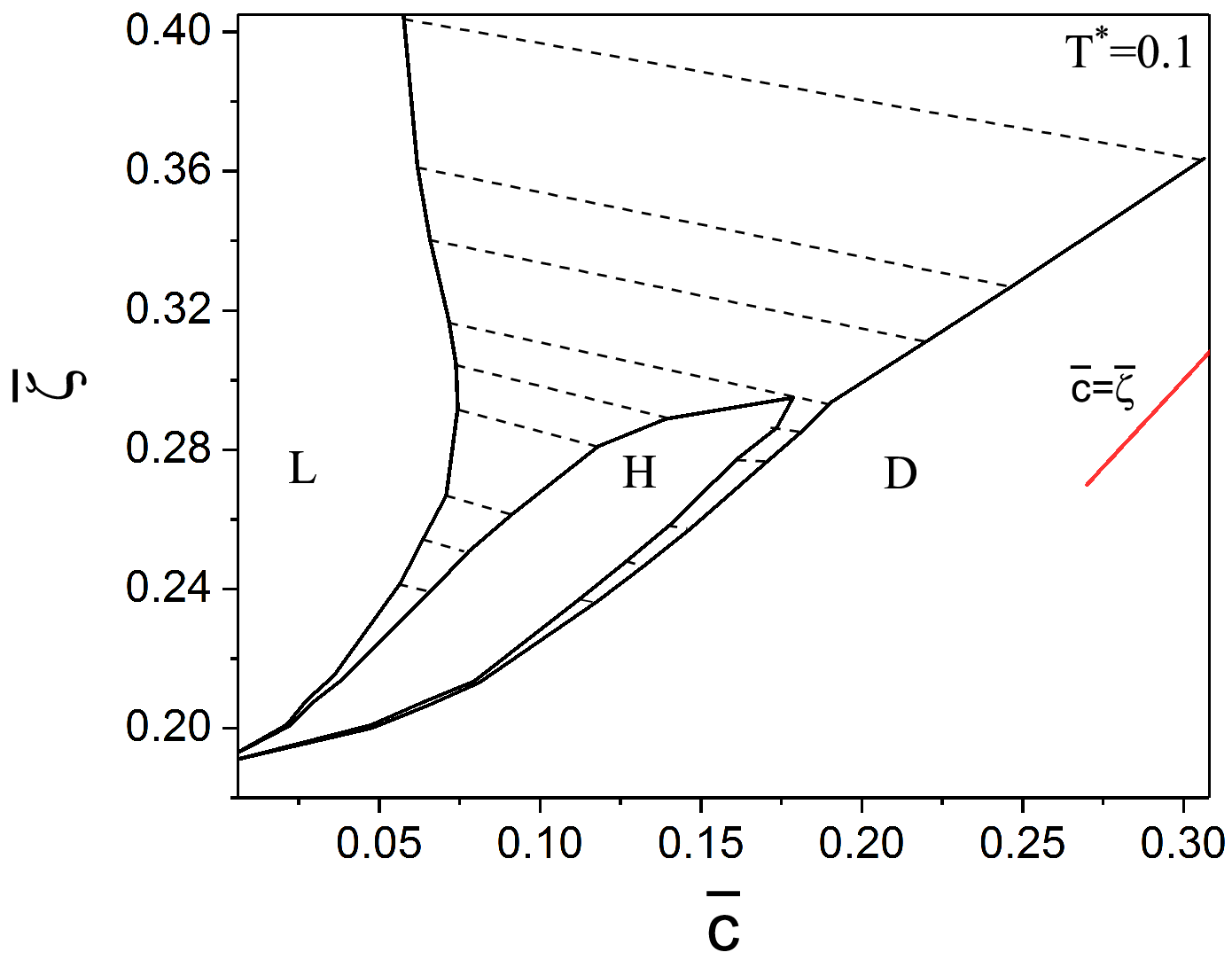}
\vskip0.5cm
\caption{The phase diagram in  the chemical potentials $(\mu_-^*,\mu_+^*)$  (left) and in concentration-volume fraction, $(\bar c,\bar\zeta)$ (right) for $T^*=0.1$. 
 L, H, D denote the lamellar (stripe), hexagonal and disordered phases, respectively. The symbols indicate  the state points for which the volume fractions $\zeta_1,\zeta_2$ are shown in Figs.\ref{fig:zetyL} and \ref{fig:zetyH}.
 Dashed lines in the right panel are the tie lines.
 In the D phase coexisting with  the H and L phases $\bar\zeta >\bar c$, i.e. D contains some amount of the second component. The short red line is $\bar c=\bar\zeta$; the region with $\bar\zeta<\bar c$  is unphysical. $T^*=k_BT/|\tilde V(k_0)|$ and  $\mu_{\pm}^*= \mu_{\pm}/|\tilde V(k_0)|$, where $\mu_{\pm}=\frac{3}{\pi}(\mu_1\pm\mu_2)$, and $\mu_i$ is the chemical potential of the i-th species. }
\label{fig:diag_T0_1}
\end{figure}

\begin{figure}
\includegraphics[scale=0.33]{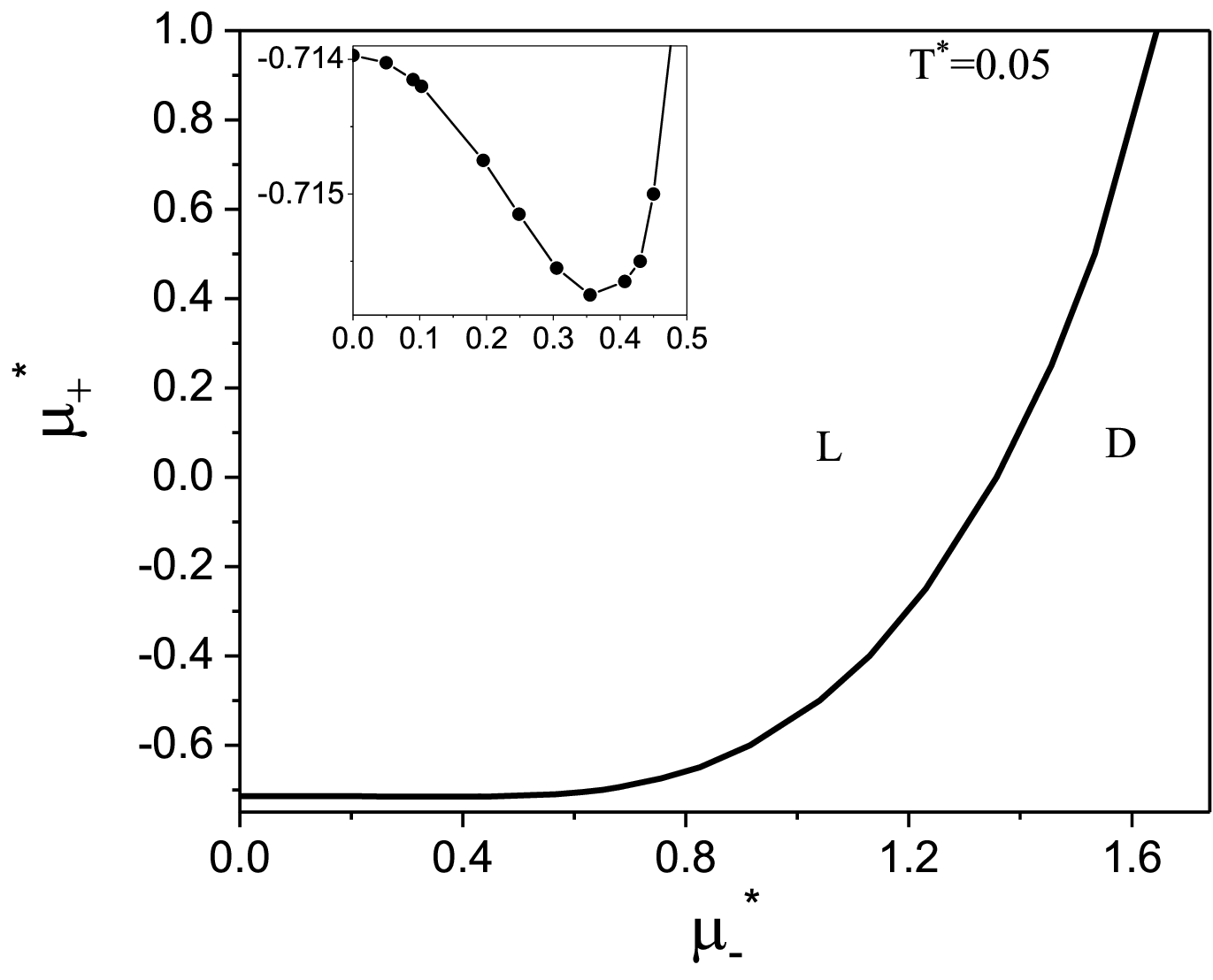}
\includegraphics[scale=0.33]{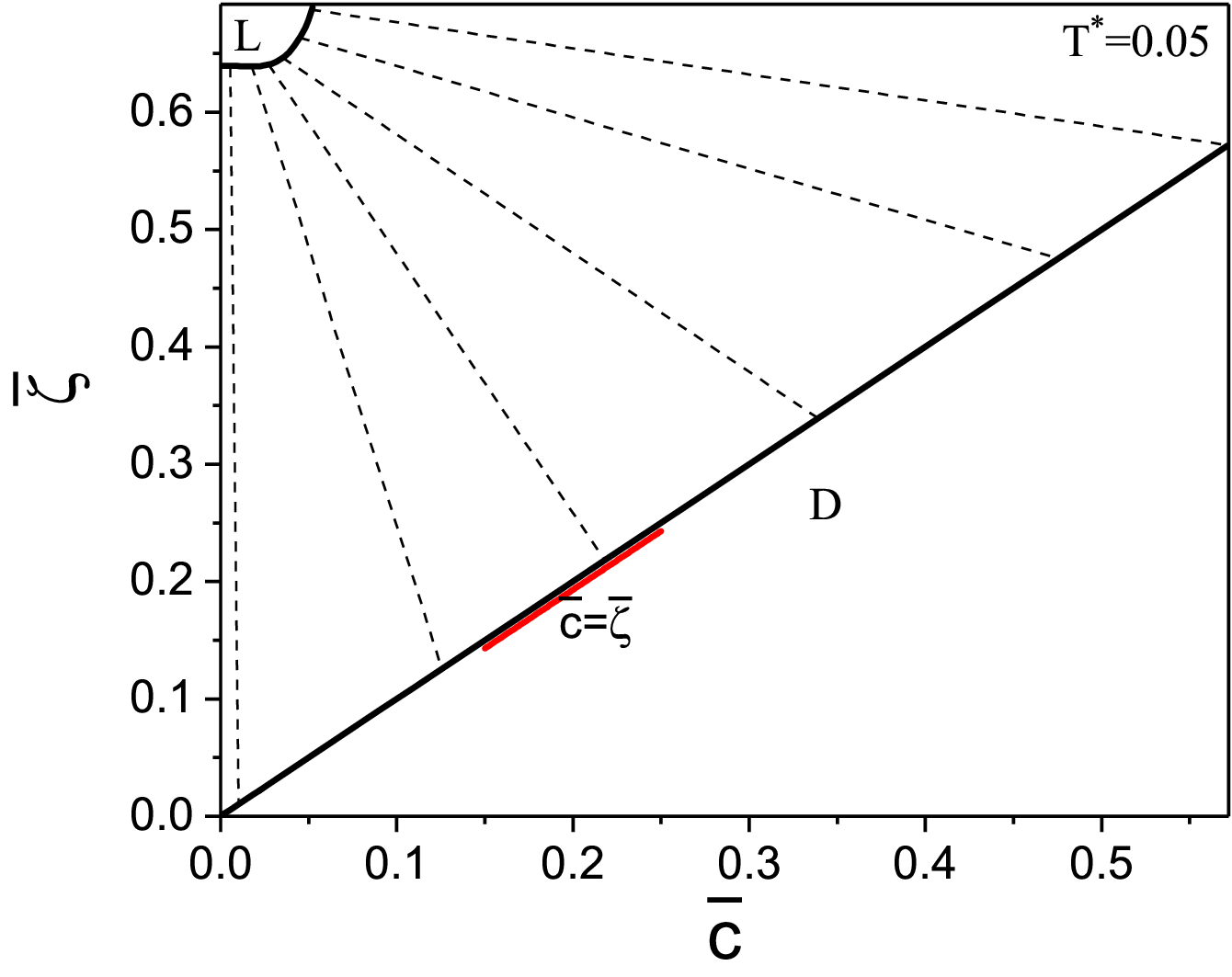}
\vskip0.5cm
\caption{The phase diagram in  the chemical potentials $(\mu_-^*,\mu_+^*)$  (left) and in the concentration-volume fraction, $(\bar c,\bar\zeta)$ (right)  for $T^*=0.05$.  Dashed lines in the right figure are the tie lines. The red bar lies on the $\bar\zeta=\bar c$ line, meaning that the L phase coexists with pure one-component fluid, except from very small $\bar c$, where $\bar\zeta>\bar c$ (invisible on the plot). The region with $\bar\zeta<\bar c$ is unphysical. $T^*=k_BT/|\tilde V(k_0)|$ and  $\mu_{\pm}^*= \mu_{\pm}/|\tilde V(k_0)|$, where $\mu_{\pm}=\frac{3}{\pi}(\mu_1\pm\mu_2)$, and $\mu_i$ is the chemical potential of the i-th species. }
\label{fig:diag_T0_05}
\end{figure}

Following the one-component case, we construct the 
 $(\bar c,T^*)$ phase diagram for fixed $\bar\zeta$ based on the minima of $\Delta\beta\omega$, which can be done  easily.  In Fig.\ref{fig:c-T_diag}, the diagram obtained by the minimization of  $\Delta\beta\omega$ with respect to $\Phi,\Psi$ for fixed $\bar c,\bar\zeta$ is presented
 for  $\bar\zeta=0.4$ and $T^*>0.08$. Importantly, this procedure can give the correct sequence of the ordered phases when the concentrations and densities in the coexisting periodic phases are similar. We verify the diagram shown in Fig.~\ref{fig:c-T_diag} by the minimization of the grand potential with fixed $T^*, \mu_+^*,\mu_-^*$. The diagrams obtained in this way are shown in Figs.~\ref{fig:diag_T0_15}, \ref{fig:diag_T0_1} and \ref{fig:diag_T0_05}, for fixed $T^*$ higher than the temperature at which the one-component system forms ordered patterns.

 As can be seen in Figs.\ref{fig:diag_T0_15} and \ref{fig:diag_T0_1}, the diagram obtained by the minimization of $\Delta\beta\omega$ and shown in Fig.\ref{fig:c-T_diag} is qualitatively correct for $T^*=0.15$, but incorrect for  $T^*=0.1$.
   For  $T^*=0.1$, the hexagonal phase  is absent at $\bar\zeta =0.4$, and instead of it, a two-phase coexistence region between the L and D phases is present. Thus, by disregarding the different values of $\bar c, \bar\zeta$ in the coexisting phases, one can obtain a wrong sequence of the stable phases. Minimization of $\beta\omega$ with fixed chemical potentials, where $\bar c$ and $\bar\zeta$ in the coexisting phases can be different, is necessary to get qualitatively correct phase diagram in the self-assembling mixture.

\begin{figure}
\includegraphics[scale=0.3]{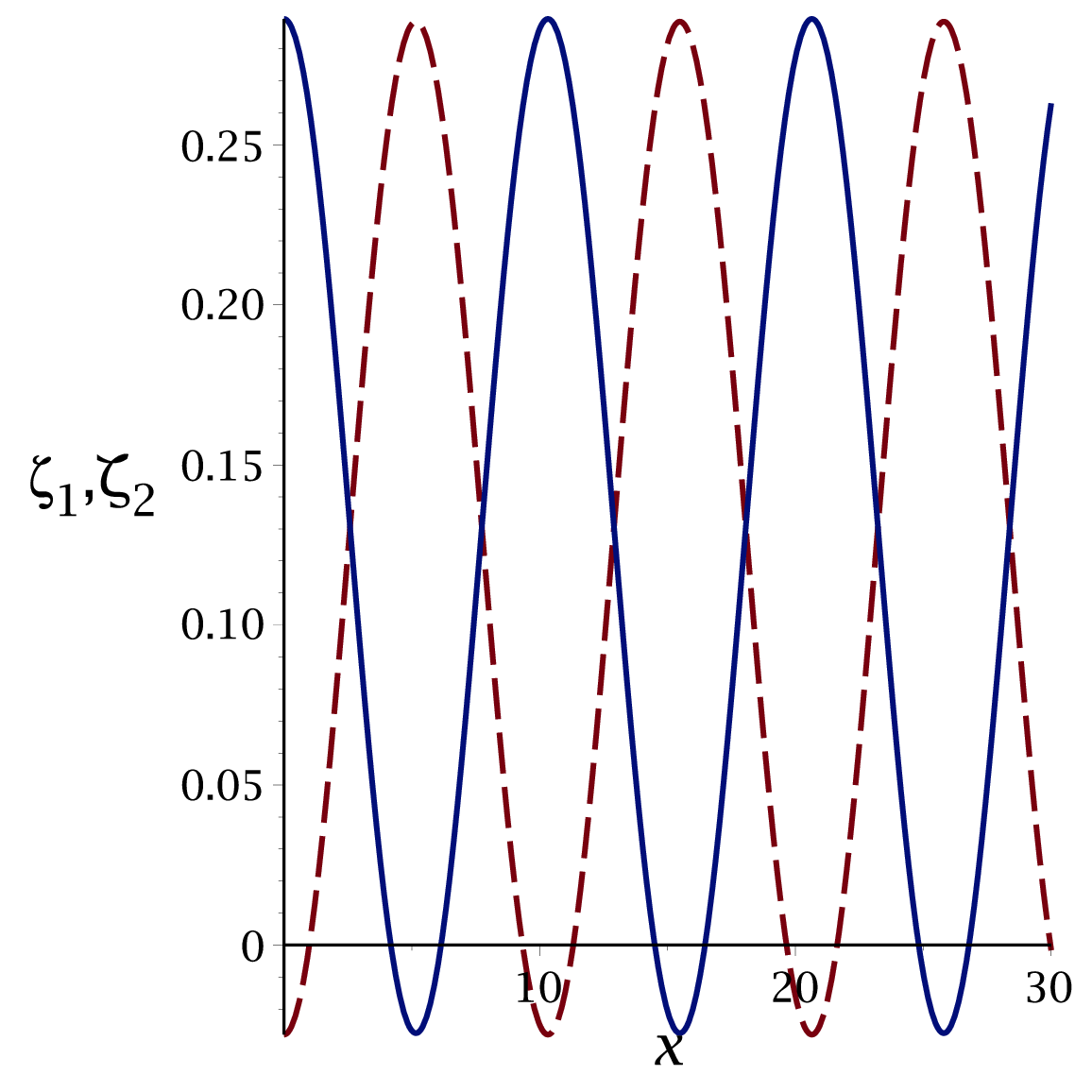}
\includegraphics[scale=0.3]{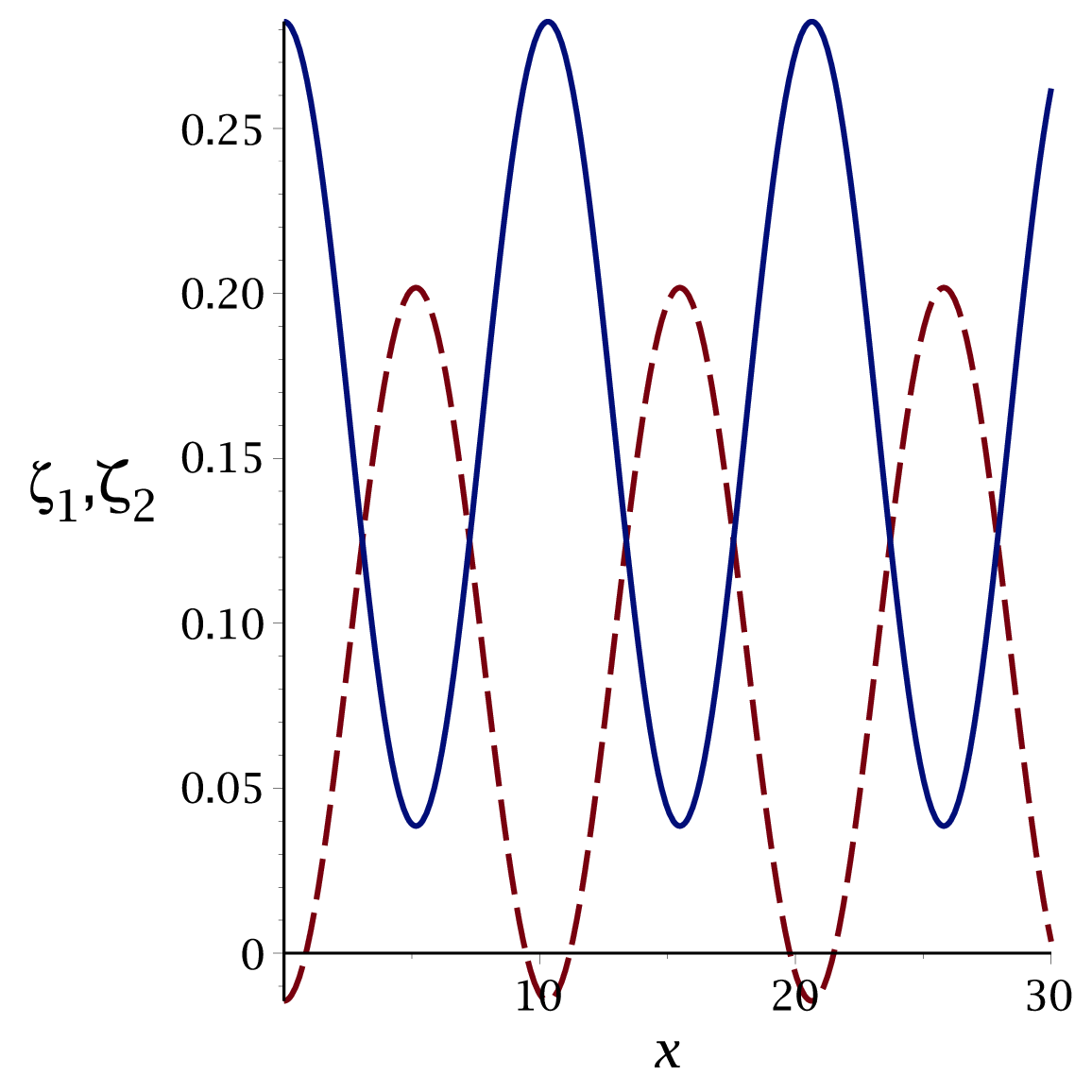}
\vskip0.5cm
\caption{The volume fractions $\zeta_1(x,0)$ (blue solid line) and $\zeta_2(x,0)$ (red dashed line)  in the L phase  in the direction $x$ of the density oscillations for $T^*=0.1$ and $\mu_+^*=0. 35$. In the left and right figure,  $ \mu_-^*=0.001$ (in the middle of the stability region of the L phase, red square in Fig.\ref{fig:diag_T0_1})  and  $ \mu_-^*= 0.08$ (at the coexistence with the H phase, green circle in Fig.\ref{fig:diag_T0_1}). 
Negative volume fraction is the artifact of the one-shell approximation.  }
\label{fig:zetyL}
\end{figure}

\begin{figure}
\includegraphics[scale=0.3]{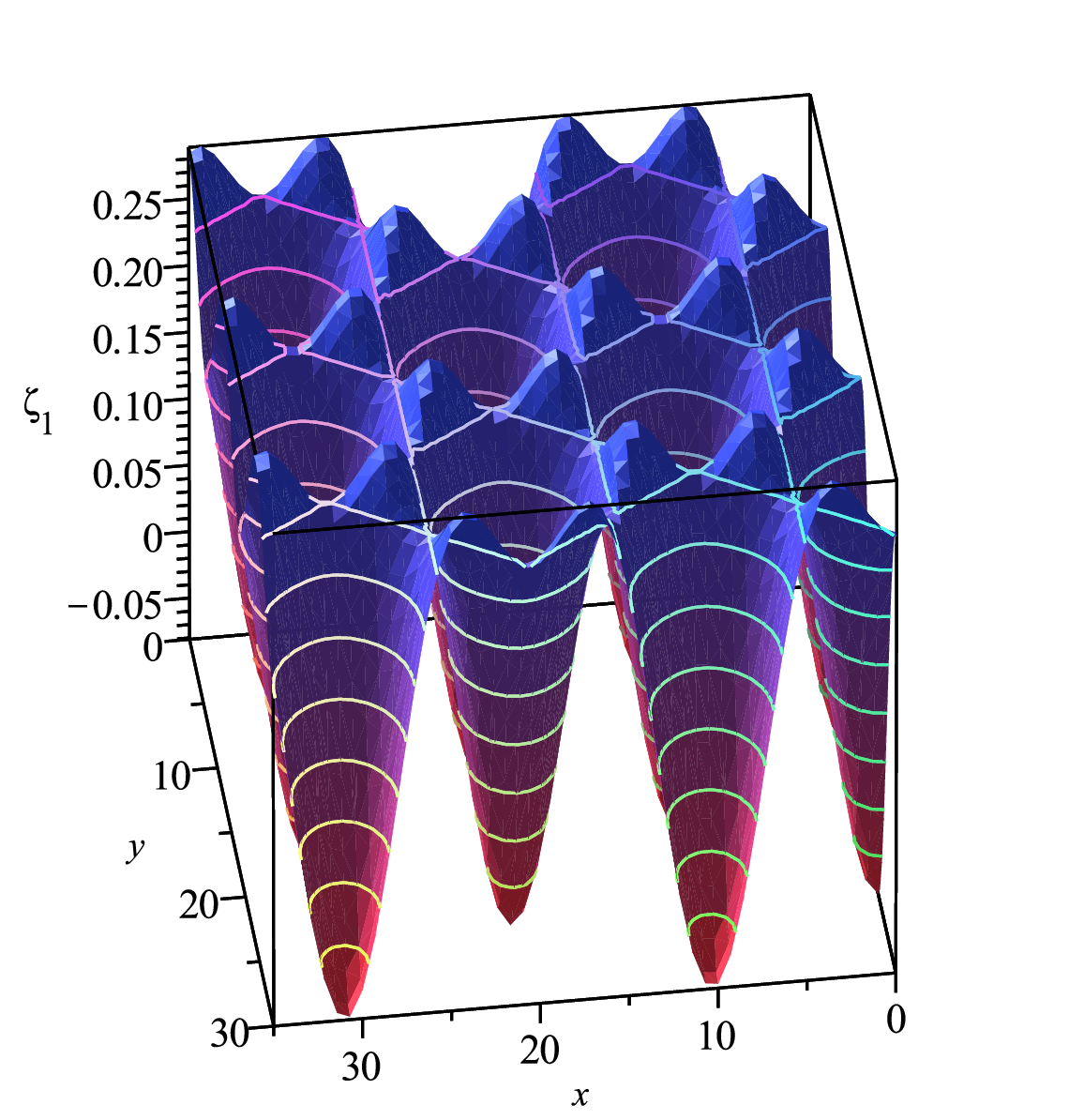}
\includegraphics[scale=0.3]{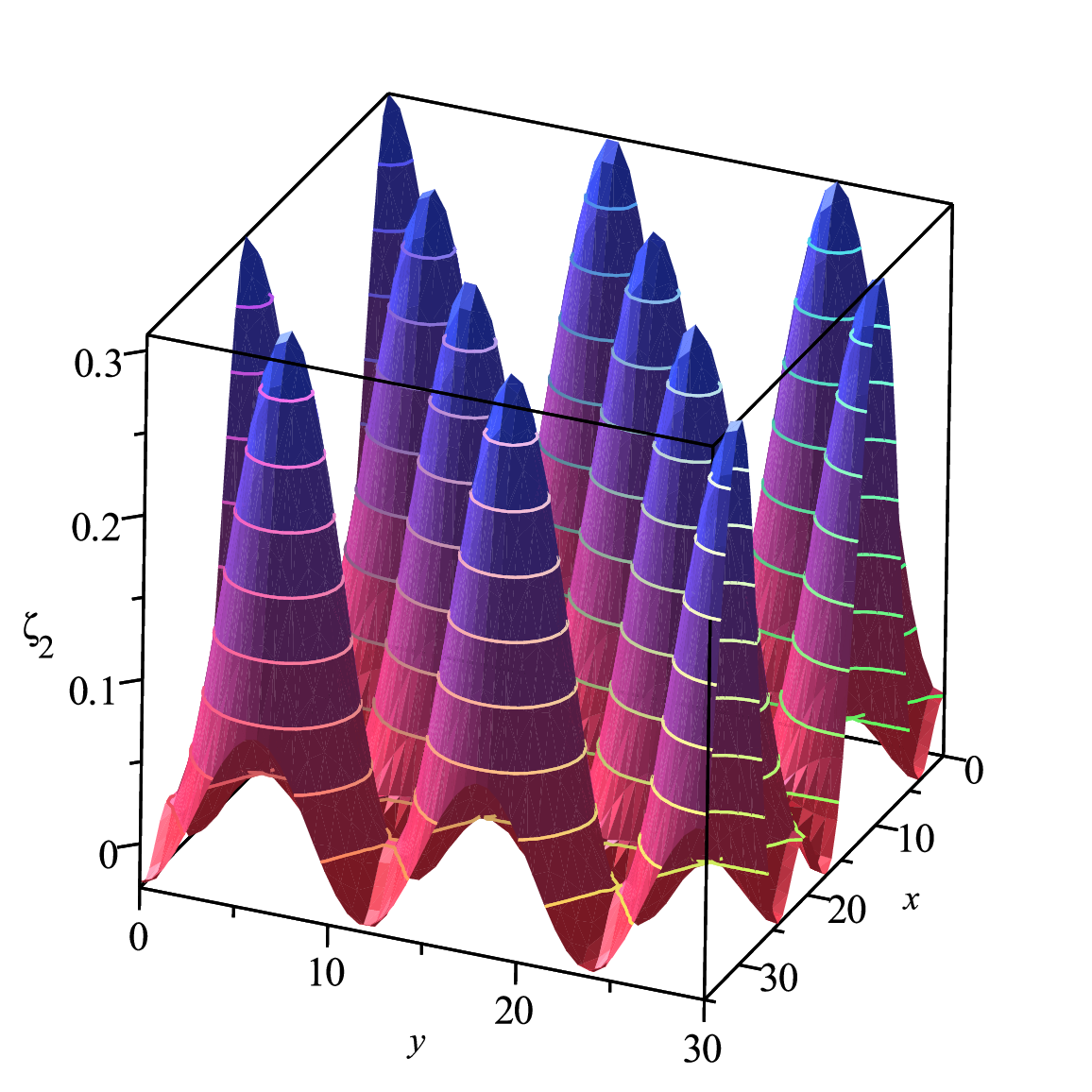}
\includegraphics[scale=0.3]{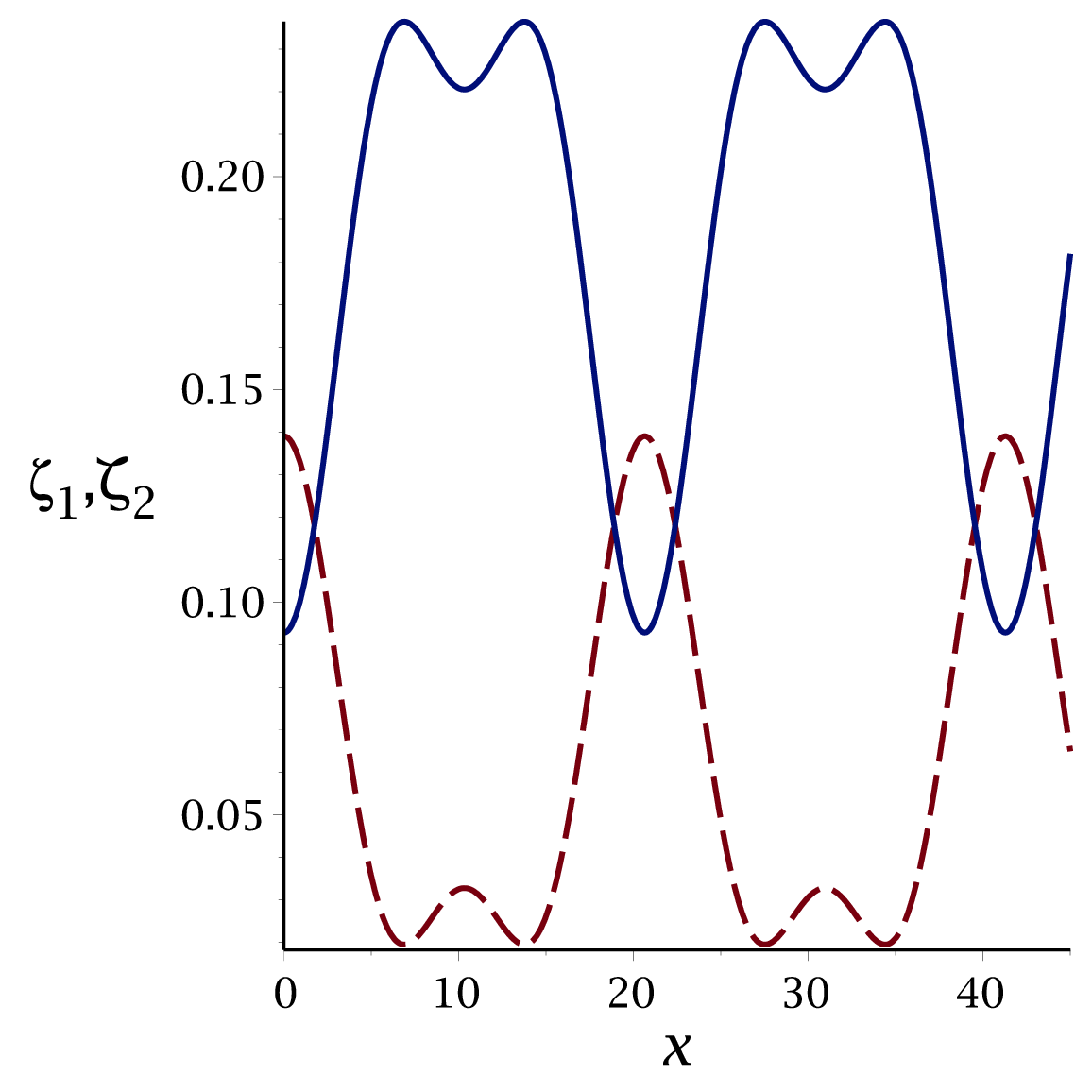}
%
\caption{The volume fractions $\zeta_1(x,y)$ and $\zeta_2(x,y)$ of the first and the second component, respectively, in the H phase. Top panels: at the coexistence with the L phase with $T^*=0.1,\mu_+^*=0.35, \mu_-^*\approx 0.08$ (green circle in Fig.\ref{fig:diag_T0_1}). Bottom panel: at the coexistence with the D phase with  $T^*=0.1,\mu_+^*\approx 0.35,\mu_-^*\approx 0.111$  (blue diamond in  Fig.\ref{fig:diag_T0_1}).   
Negative volume fraction is the artifact of the one-shell approximation.  
In the bottom panel,  $\zeta_1(x,0)$ (blue solid line) and $\zeta_2(x,0)$ (red dashed line) are shown in the same plot for comparison.
}
\label{fig:zetyH}
\end{figure}

For illustration of the structure of the L and H phases, we choose 
  $T^*=0.1$, $\mu_+^*= 0.35$, and several values of $\mu_-^*$, namely $\mu_-^*=0.001$ inside the stability region of the L phase,  $\mu_-^*=0.08$ at the L-H  phase coexistence  and $\mu_-^*=0.111$ at the H-D  phase coexistence. 
 The volume fractions of the two components in the L and H phases at the above state points are shown in Figs.\ref{fig:zetyL} and \ref{fig:zetyH}.
In both phases, $|\Phi|\gg |\Psi|$, i.e. mainly the concentration oscillates. Moreover, in the L-phase, $\bar c$ is quite small, meaning rather small asymmetry between the stripes rich in the first and the second component even at the coexistence with the H or D phases, where the largest asymmetry between the stripes takes place (see Fig.~\ref{fig:zetyL}).

\begin{figure}
\includegraphics[scale=0.33]{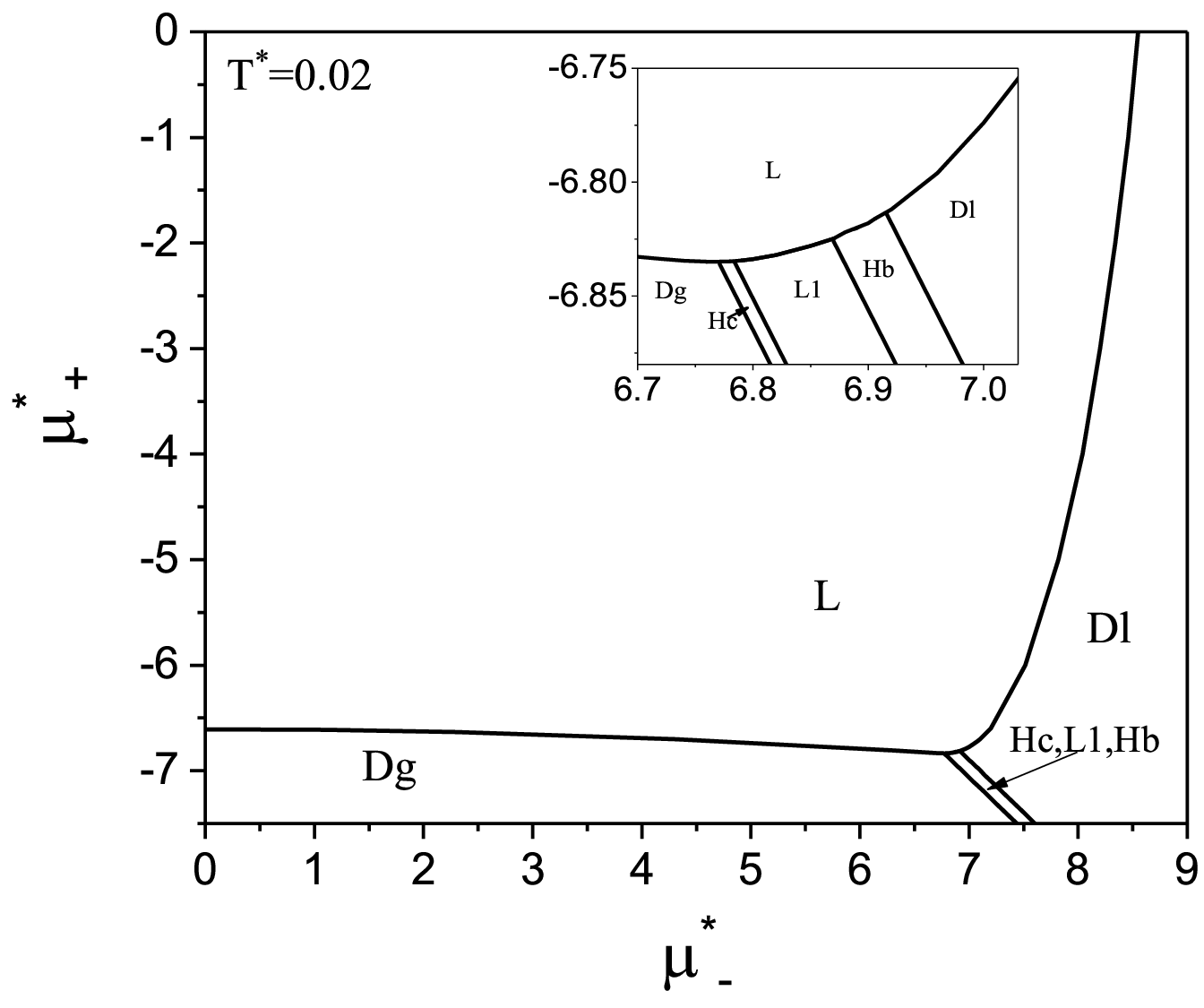}
\includegraphics[scale=0.33]{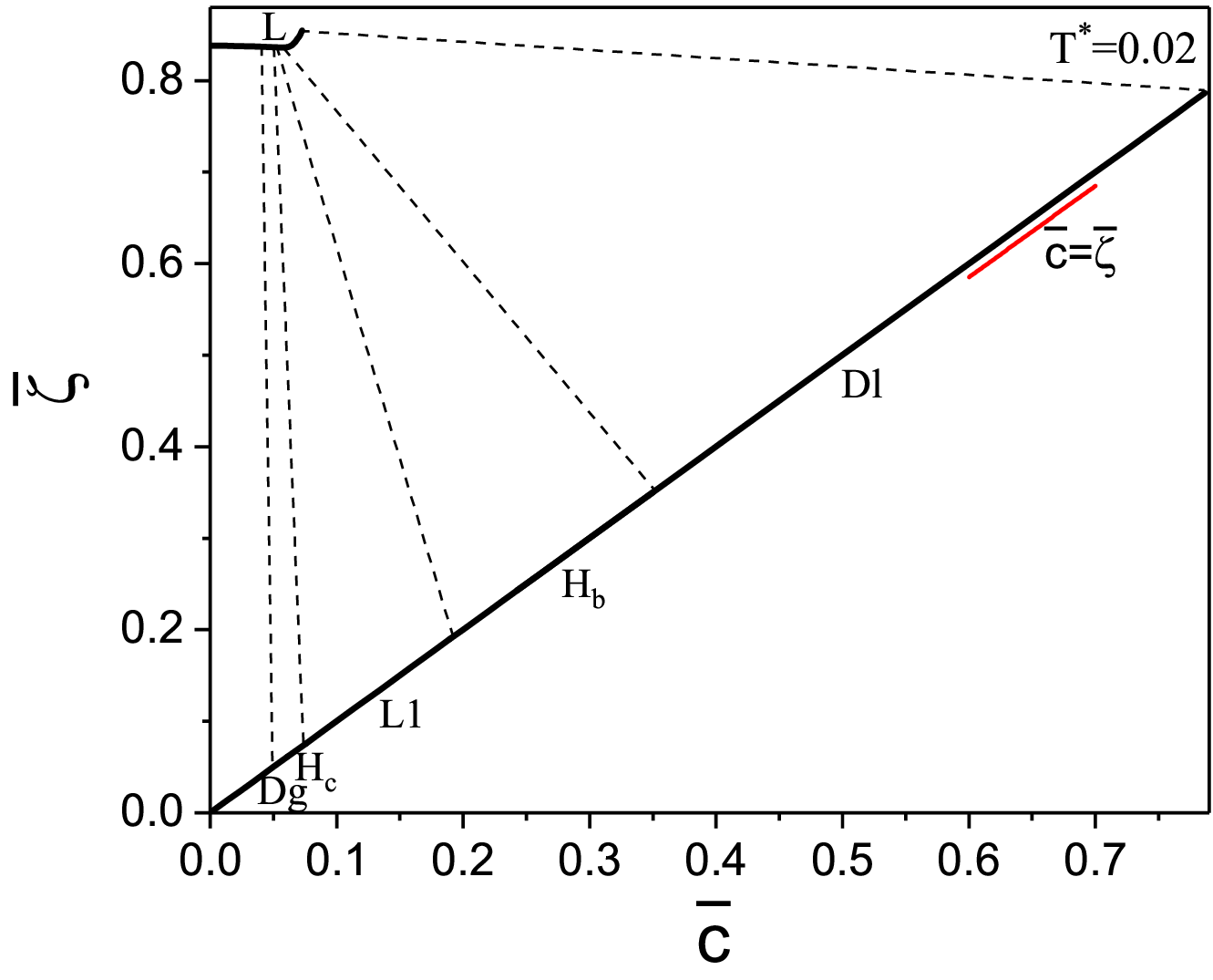}
\vskip0.5cm
\caption{The phase diagram in  the chemical potentials plane $(\mu_-^*,\mu_+^*)$ (left) and in the $(\bar c,\bar\zeta)$ plane (right) for $T^*=0.02$. The L phase with parallel stripes of alternating components coexists with the disordered gas (Dg) and the disordered liquid rich in the first component (Dl), and with periodic phases in the  one-component system between the two triple points L-Dg-Hc and L-Hb-Dl, where Hc and Hb denote the hexagonal pattern of clusters and bubbles, respectively, and L1 denotes the one-component lamellar phase with  stripes separated by empty layers. In the inset in the left panel, the bottom-right corner where the ordered phases in the  one-component system are stable is expanded. Dashed lines in the right panel are the tie lines. The $\mu_-^*<0$ part of the diagram is not shown, since $\mu_+^*$ is the symmetry axis. $T^*=k_BT/|\tilde V(k_0)|$ and  $\mu_{\pm}^*= \mu_{\pm}/|\tilde V(k_0)|$, where $\mu_{\pm}=\frac{3}{\pi}(\mu_1\pm\mu_2)$, and $\mu_i$ is the chemical potential of the i-th species. }
\label{fig:diag_T_0_02}
\end{figure}

The diagram for temperature low enough for ordering of the one-component system is shown in Fig.\ref{fig:diag_T_0_02} for $T^*=0.02$. We can see that the L phase with oscillating concentration coexists with very dilute gas (almost vacuum) for $ \mu_+^*<-6$, and with the liquid of the first component for 
$\mu_-^*>7$. In the corner with $\mu_+^*\approx -6.8,\mu_-^*\approx 6.8$, the L phase with oscillating concentration coexists with the one-component ordered phases stable between the parallel lines $\mu_+^*\approx \mu_{tr}^*-\mu_-^*$, where $\mu_{tr}^*$ is the chemical potential of the first component at the coexistence between the one-component ordered phases.
We verified that for the considered temperatures, the chess-board pattern was only metastable.

  \subsection{The simulation results}
  \label{MC}
  
To verify and illustrate our predictions, we performed MC simulations in the $\mu$VT ensemble. The purpose of this study was to check if the  patterns obtained in our MF theory indeed occur in some part of the phase space, and if these are the only patterns that can be present in the considered mixture.  For this purpose we chose hard spheres with the same diameter $a$ that sets the length unit, $u_{11}(r)=u_{22}(r)=u(r)=-u_{12}(r)$ and the double Yukawa potential  (\ref{Yukawa}) for $u$.  The cut-off radius was 15.

In the simulations, the particles can be displaced, destroyed or
created. The acceptance probability for both these moves is determined by the Metropolis criterion $\min[1, \exp(- \beta \Delta U)]$, where $\Delta U$ is the change of the total energy due to a trial move. In a creation trial move a new molecule is created at a random location in  the simulation box with a volume $V_s$, with the acceptance probability $\min[1, z_{i} \exp(- \beta \Delta U) V_s/(N_{i} + 1)]$, where $z_{i}$ is the configurational activity of species $i$ defined by $z_{i}=\exp(- \beta \mu_{i})/\hat I_{i}^{3}$ and $\hat I_{i}$ is the de Broglie thermal wavelength; in a deletion trial move a molecule is randomly chosen and deleted from the system with the acceptance probability $\min[1, N_{i} \exp(- \beta \Delta U)/(z_{i} V_s)]$ \cite{allen2017computer, frenkel2023understanding}.

In order to simulate quasi-2D systems such as inclusions in biological membranes, we consider a simulation box with the edges $L_{x}, L_{y}, L_{z}$, with $L_{x}=L_{y} = 80a$, (in some cases $L_{x}=L_{y} = 45a$ to speed up the simulations), and applying  periodical boundary conditions with the minimum image convention in these two directions. In the $z$ direction, however, the particles were confined by impenetrable hard walls and $L_{z}=1.5 a$. Each system has run $10^{6}$ MC steps for equilibration and 
 $5\times 10^5$ for production.

We performed simulations along several paths in the phase space $(\mu_1^*,\mu_2^*,T^*)$. In the first family of paths, $\mu_1^*=\mu_2^*$, and in the second one  $\mu_1^*$ was fixed, and $\mu_2^*$ varied. The obtained patterns generally agree with our theoretical predictions. In particular, for same region of the chemical potentials, the clusters of the minority component appear at higher $T^*$ than the symmetrical stripes. 

\begin{figure}
\includegraphics[scale=0.38]{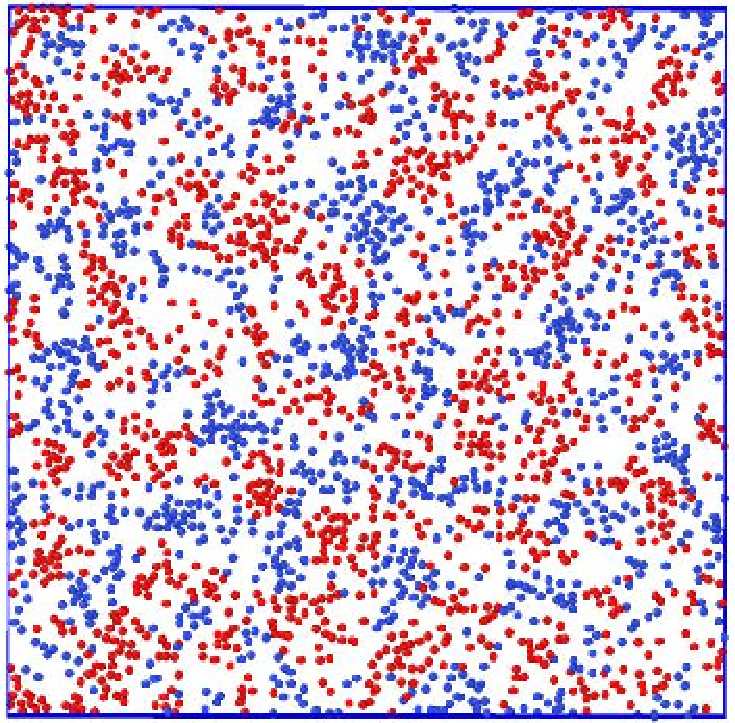} \quad
\includegraphics[scale=0.38]{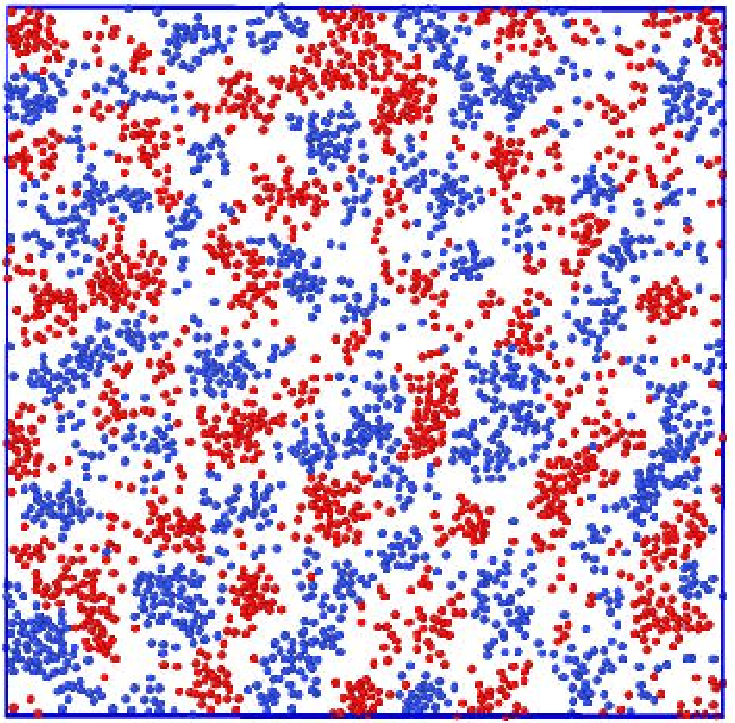} \quad
\includegraphics[scale=0.38]{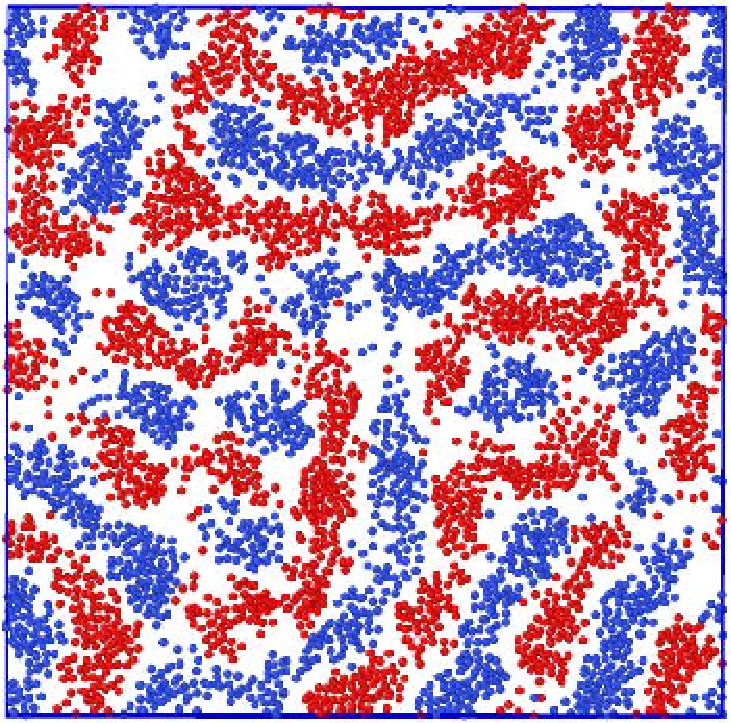}
\caption{Representative  snapshots for equal chemical potentials  $ \mu_1^*=\mu_2^* =\mu_+^*= -0.095$ showing the inhomogeneous disordered structure for increasing density. From the left to the right figure, $ T^*=0.03,0.023,0.021$, respectively. Blue and red circles represent the first and the second component particles, respectively.}
\label{fig:snapDs}
\end{figure}

\begin{figure}
\includegraphics[scale=0.33,angle=90]{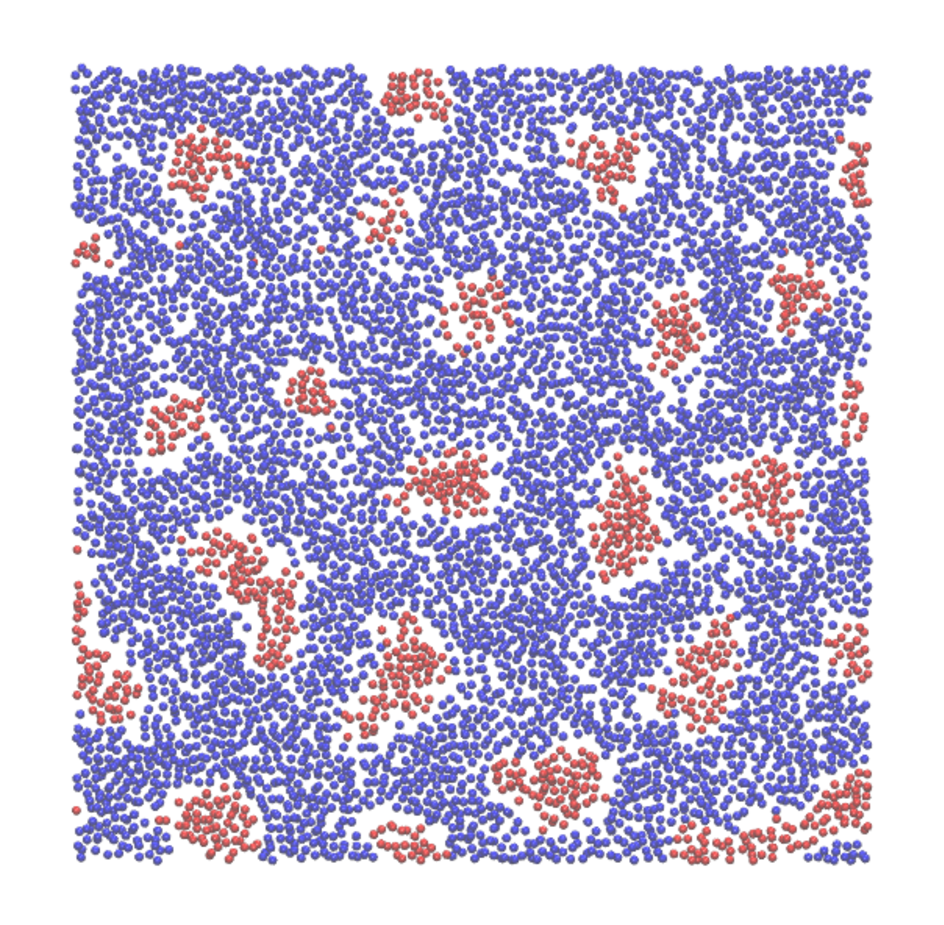}
\includegraphics[scale=0.33,angle=90]{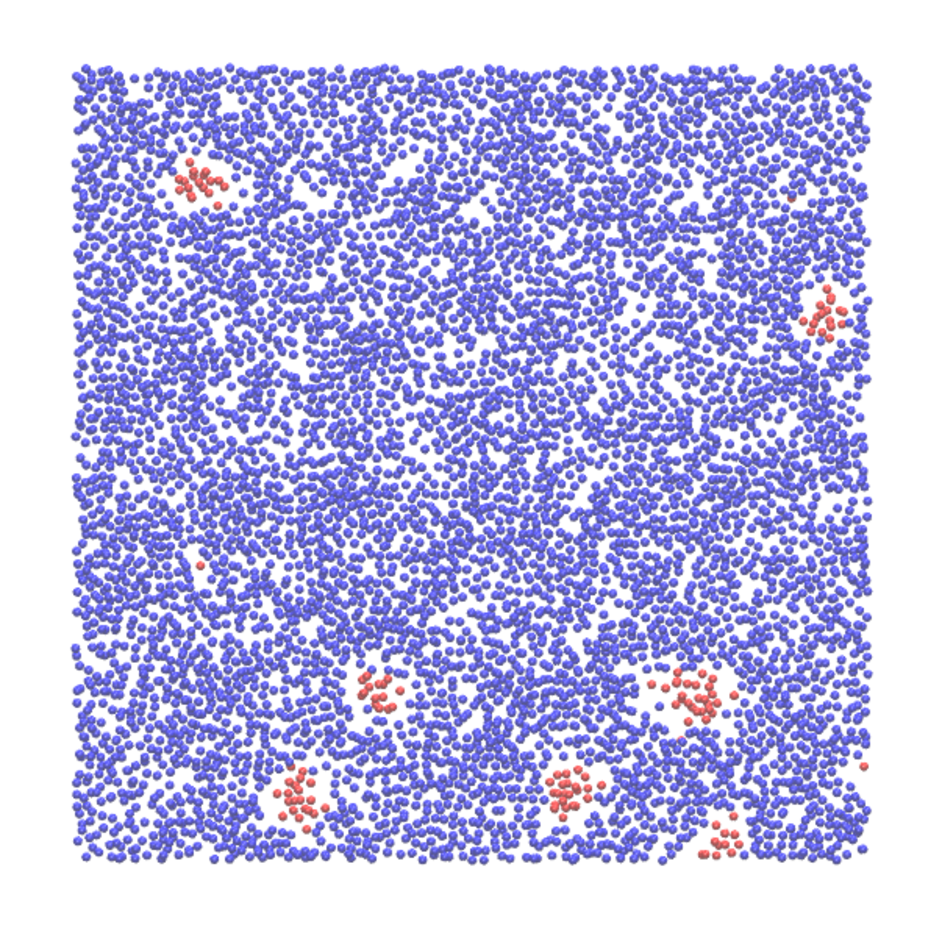}
\includegraphics[scale=0.33,angle=90]{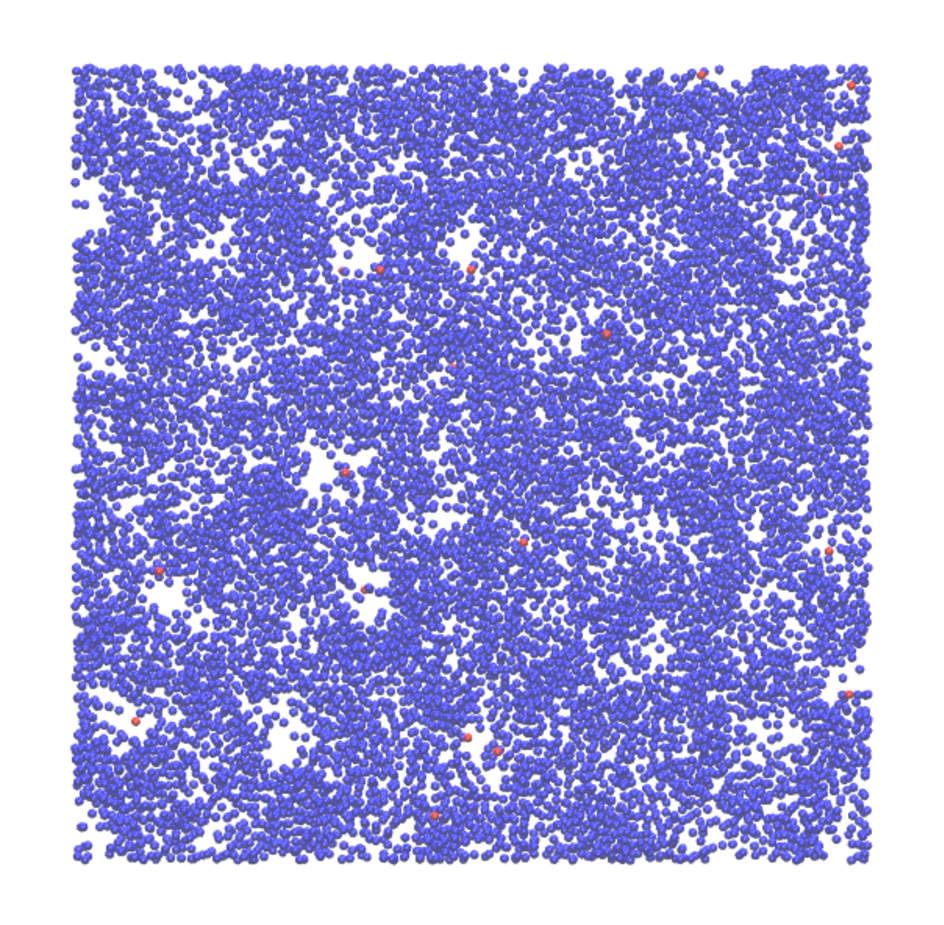}
\caption{Representative snapshots for different chemical potentials showing the liquid rich in the first component for $ T^*=0.028$. From the left to the right figure, $\mu_1^*=-0.019,-0.019,-0.076$ and $\mu_2^* =-0.057, -0.076,-0.265$, respectively. Blue and red circles represent the first and the second component particles, respectively.}
\label{fig:snapDa}
\end{figure}

\begin{figure}
\includegraphics[scale=0.4]{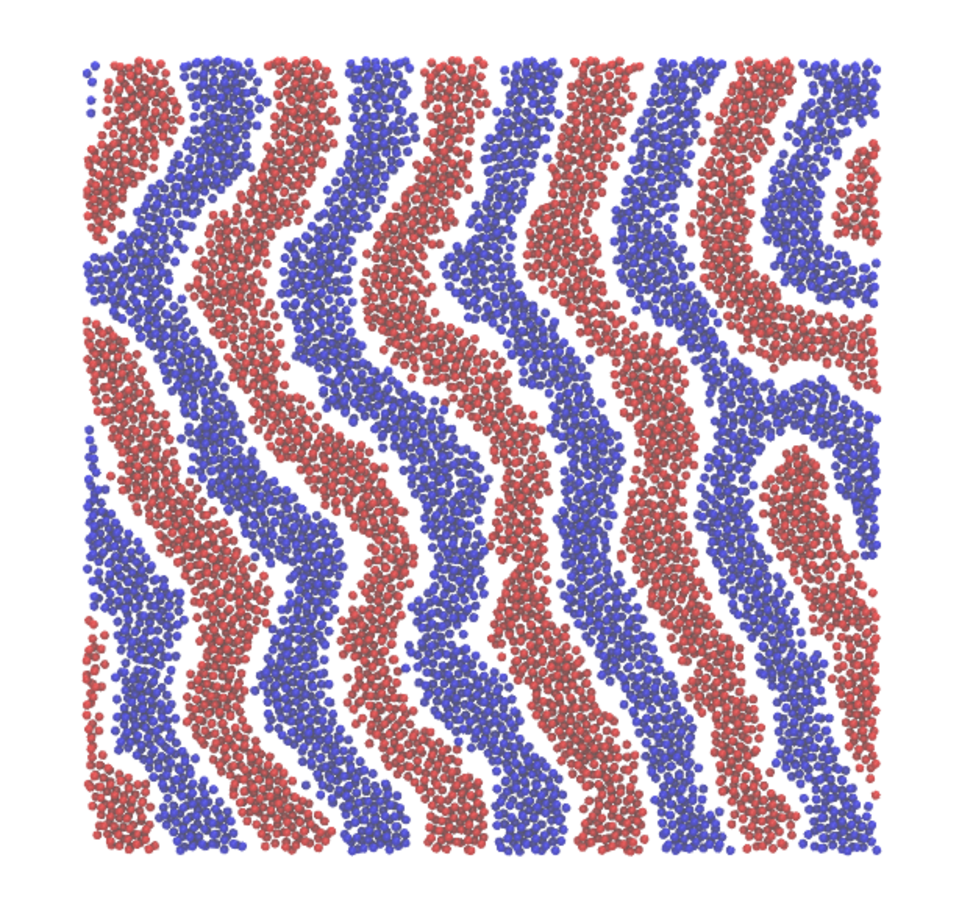}
\includegraphics[scale=0.4]{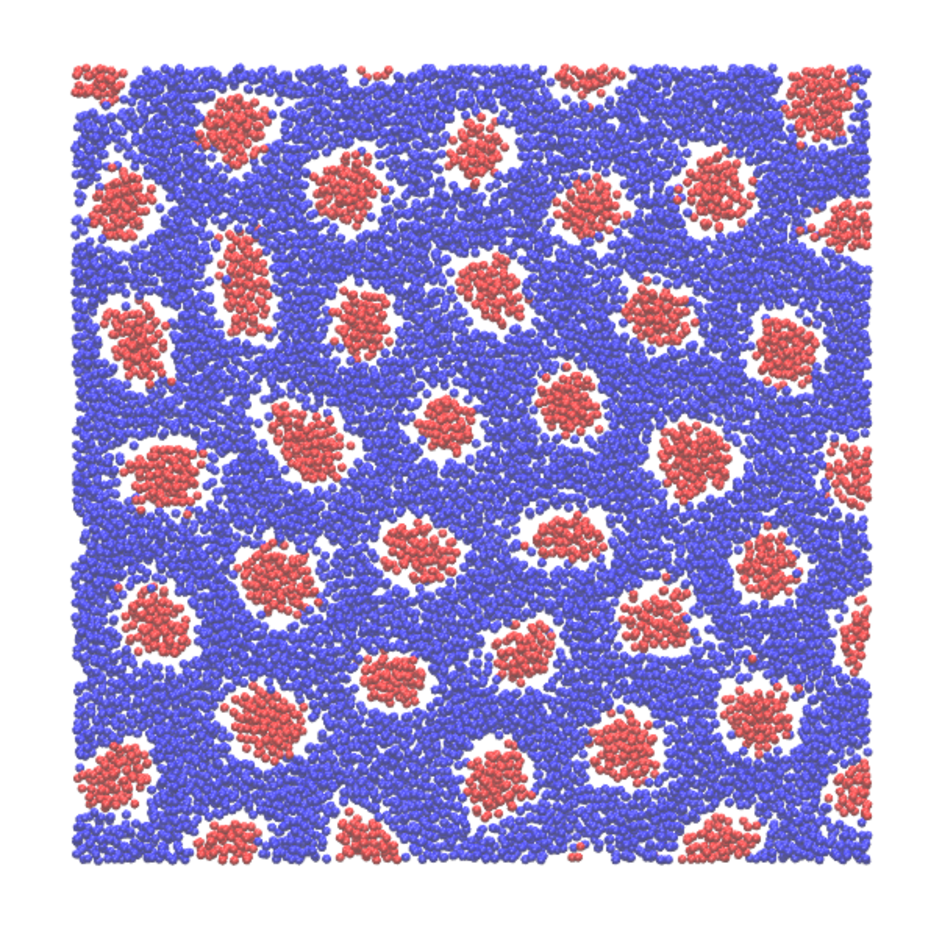}
\caption{Representative  snapshots of dense patterned structures. Left panel:  $T^*=0.015$, $\mu_1^*=\mu_2^*=\mu_{+}^* =-0.11$. Right panel: $T^*=0.028$, $\mu_1^*=-0.076$ and $\mu_2^* =-0.17$ ($\mu_{+}^*=-0.123$, $\mu_{-}^*=0.047$). Blue and red circles represent the first and the second component particles, respectively.}
\label{fig:snapO}
\end{figure}

Representative snapshots are shown in Figs.\ref{fig:snapDs}-\ref{fig:snapO}. In Fig.\ref{fig:snapDs} we show the disordered phase with mixed components, whereas in Fig.\ref{fig:snapDa} the disordered liquid rich in the first component is shown. In the symmetrical mixture, we show how the structure evolves for increasing density, i.e.  for decreasing $T^*$. Note the increasing inhomogeneity of the concentration for increasing density. 
 For $\bar c>0$, self-assembly of the minority component into clusters occupying  the voids present in the majority component is clearly seen in Fig.\ref{fig:snapDa}. The number and size of the clusters decreases for increasing  $\mu_-^*=\mu_1^*-\mu_2^*$, and formation of the pure one-component liquid can be observed.

In Fig.\ref{fig:snapO}, we show  the alternating stripes of the two components for $\bar c=0$, and clusters of the minority component for $\bar c>0$. 
The stripes can be interpreted as the L phase with some defects that are typically present in experimental lamellar phase as well. The clusters in Fig.\ref{fig:snapO} show hexagonal order with some displacements of the clusters. This snapshot (as well as several other ones) suggests that the ordered H phase can occur in the true equilibrium state in this region of the phase diagram on the level of ensemble-averaged densities.  We did not observe the ordered chess-board structure, except from local chess-board patterns at relatively high temperature (Fig.\ref{fig:snapDs}, central panel).

\section{Discussion and summary}
\label{sec:discuss}

We have found that when two types of SALR particles are mixed together and the 
cross-interaction is repulsive at short and attractive at large separations, then ordered patterns can be formed up to much higher temperature than in the one-component system.   This conclusion concerns  the class of models with the interactions $u_{12}(r)=-u_{ii}(r)$, such that $V(r)=(\frac{6}{\pi})^2u_{ii}(r)\theta(r-1)$ takes in Fourier representation a global minimum for $k_0>0$. According to our MF stability analysis, the disordered phase in the one-component system looses stability with respect to periodic ordering at the largest temperature $T_{\lambda}^*(\bar \zeta,\bar \zeta)\approx 0.0247$ for $\bar\zeta\approx 0.15$. For the same  density of each component in the mixture (for $\bar c=0$), the temperature at the instability of the disordered phase is $T_{\lambda}^*(0,0.3)\approx 0.16$, i.e. it is much higher. Our $\lambda$-surface depends on the interaction potential $V$ only through the temperature scale  set by  $|\tilde V(k_0)|$. Based on this universality, we conclude that it is much easier to obtain self-assembled ordered structures in a mixture with the two mermaids and a peacock interactions, than in the absence of the second component. 

In our theory, apart from the temperature and chemical potentials scale set by  $|\tilde V(k_0)|$, the  phase diagram depends on the interactions only through the single parameter $\tilde V(0)$ (see Eq.(\ref{boXZ})). This means that our results, when expressed in terms of dimensionless variables, should concern all versions of the two mermaids and a peacock model, in which the integral of $V(r)$ takes the same value. 
We considered interactions such that  $\tilde V(0)=-0.363<0$ ($\tilde V(0)/|\tilde V(k_0)|\approx-0.036$), leading to rather large clusters. 

   For equal chemical potentials, alternating stripes of the first and the second component of the width $\pi/k_0$ were obtained in the previous theoretical and simulation study of a 3D system~\cite{patsahan:21:0}. The  natural question was how this structure evolves upon changing the chemical potentials or temperature.
   Our results show that in 2D the phase with alternating stripes of the two components is stable only up to quite small density difference of the two components, $|\bar c|\ll \bar\zeta$, for all temperatures studied (Figs.\ref{fig:diag_T0_15}-\ref{fig:diag_T0_05} and \ref{fig:diag_T_0_02}). 
When the density in the stripes of the second component decreases to about $0.7$ times the density in the stripes of the majority component, a first-order transition to a phase poor in the second component takes place (see Fig.~\ref{fig:zetyL} for $\zeta_i$ at the  L-H phase coexistence).

At low $T^*$, the L phase is very dense and  occupies a small region in the $(\bar c,\bar \zeta)$ diagram (see Figs.\ref{fig:diag_T_0_02} and \ref{fig:diag_T0_05}). 
The density in the L phase at the coexistence with the dilute gas  decreases from very large values to rather small values upon increasing $T^*$ (see Figs.\ref{fig:diag_T0_15} and \ref{fig:diag_T0_1}). 

One could expect that upon increasing the chemical potentials difference, a transition from the L phase to the H phase with clusters of the minority component in the liquid of the majority component would occur, because of relatively similar $\bar c$ and $\bar\zeta$  in the two phases.  We found out, however that at low $T^*$, the L phase coexists with the phase rich in the first component, either disordered for large $\mu_+^*$, or ordered for $\mu_-^*\approx -\mu_+^*$. This is because the H phase appears only at relatively large $T^*$, and its stability  region on the $(\mu_-^*,\mu_+^*)$ plane increases with increasing $T^*$. We should stress that the jump of $\bar c$ or $\bar\zeta$ in the coexisting L and Dg, Hc, L1, Hb, Dl phases is very large, and that the density in the Hc, L1, Hb phases is significantly smaller than the density of the L phase coexisting with them (see Fig.\ref{fig:diag_T_0_02}).

In principle one could expect also  chess-board pattern (see Fig.\ref{fig:patterns}, central panel) or chains of particles separated by empty spaces, observed in Ref.\cite{virgiliis:23:0} for the lattice version of the model but with $\tilde V(0)>0$. We did not see such ordered patterns in MC simulations of our model (\ref{Yukawa}), except from the local chess-board pattern shown in Fig.~\ref{fig:snapDs}. In the theory, we found that the chess-board structure was metastable. 

  We should note that in the model, the interaction potentials are fixed, i.e. depend only on the distance between the particles, whereas in particular systems, the interactions may depend on the thermodynamic state.
Our results, however, shed light on the possible ordered phases, and on the coexistence between them.

Let us finally comment on the validity of the phenomenological Landau-Brazovskii (LB) functional \cite{brazovskii:75:0} that correctly predicts the sequence of the ordered phases for the SALR particles and for the block copolymers, where the order parameter was identified with the excess density and concentration, respectively.  The same sequence of phases for increasing $\bar c$ would be predicted by the LB functional for the two mermaids and a peacock model in the case of fixed density. We have found, however that for our model the LB functional can be valid only for relatively high temperatures (see Figs.~\ref{fig:c-T_diag}--\ref{fig:diag_T0_05}). For lower temperatures, however, predictions of the LB functional are qualitatively incorrect. It is necessary to take into account different densities and concentrations, and require equal chemical potentials and pressure in the coexisting phases, as done in this work.
 
We conclude that although the ordered patterns are relatively simple, the phase diagram in the considered mixture is quite complex. We expect that the topology of the phase diagram will remain the same for various shapes of the interactions having the property $\tilde V(0)<0$.  The question how the phase diagram changes when the long-range repulsion between like particles is strong, such that $\tilde V(0)>0$ and small aggregates are formed, will be a subject of our future studies. 

\section*{Acknowledgments}
We gratefully acknowledge the financial support from the European Union Horizon 2020 research 
and innovation programme under the Marie
Sk\l{}odowska-Curie grant agreement No 734276 (CONIN). OP received the financial support from the Ministry of Education and Science of Ukraine for the implementation of the joint Ukrainian-Polish project (grant agreement No.~M/43-2023).

\section{Appendix}
\subsection{the $g$ functions and the geometric factors}
In the lamellar, hexagonal and chess-board phases the $g(x,y)$ functions take the following forms, respectively:
\begin{eqnarray}
\label{lam}
g^{L}({\bf r})=\sqrt 2\cos(k_b x)
\end{eqnarray}
\begin{eqnarray}
\label{hex}
g^{H}({\bf r})=\sqrt{\frac{2}{3}}\Bigg[\cos(k_bx)+
2\cos\Big(\frac{k_bx}{2}\Big)\cos\Big(\frac{\sqrt 3 k_by}{2}\Big)\Bigg]
\end{eqnarray}
\begin{equation}
g^{chess}({\bf r})=\cos(k_b x)+\cos(k_b y)
\end{equation}
The corresponding geometric factors are:
\begin{eqnarray}
 \label{kappanml}
\kappa_{3}^{L}=0,\hskip1cm
 \kappa_{4}^{L}=\frac{3}{2}
\end{eqnarray}
\begin{eqnarray}
 \label{kappanmh}
\kappa_{3}^{H}=\sqrt{\frac{2}{3}},\hskip1cm
 \kappa_{4}^{H}=\frac{5}{2}
\end{eqnarray}

\begin{equation}
\kappa_{3}^{chess}=0,\hskip1cm
 \kappa_{4}^{chess}=\frac{9}{4}
\end{equation}
\subsection{Expressions for $\bar c,\bar\zeta,\Phi,\Psi$ in our MF approximation}
In the  binary mixture, the minimization of the grand potential   per unit area, Eq.(\ref{bo}), with respect to $\bar\zeta$,  $\bar c$, $\Phi$ and $\Psi$ gives in the $\varphi^4$ MF approximation

\begin{eqnarray}
\label{Dbmu+}
\beta\mu_+=\beta\mu_+^D+\frac{A_3}{2}\Psi^2+\frac{\kappa_{3}A_4}{3!}\Psi^3+\frac{\kappa_{4}A_5}{4!}\Psi^4
\nonumber
 \\
-\frac{3}{2\pi}\Big(X^2+Z^2\Big)+\frac{\kappa_3}{\pi}\Big(X^3+Z^3\Big)-\frac{3\kappa_4}{4\pi}\Big(X^4+Z^4\Big),
\end{eqnarray}

\begin{eqnarray}
\label{Dbmu-}
\beta\mu_-=\beta\mu_-^D-\frac{3}{2\pi}\Big(X^2-Z^2\Big)+\frac{\kappa_3}{\pi}(X^3-Z^3)-\frac{3\kappa_4}{4\pi}\Big(X^4-Z^4\Big),
\end{eqnarray}

\begin{eqnarray}
\label{A+}
A_2\Psi+\frac{\kappa_{3}A_3}{2}\Psi^2+\frac{\kappa_{4}A_4}{3!}\Psi^3
+\frac{3}{\pi}\Big(X+Z\Big)-\frac{3\kappa_3}{2\pi}\Big(X^2+Z^2\Big)-\frac{\kappa_4}{\pi}\Big(X^3+Z^3\Big)=0
\end{eqnarray}
and
\begin{eqnarray}
\label{A-}
-\frac{1}{T^*}\Phi
+\frac{3}{\pi}\Big(X-Z\Big)-\frac{3\kappa_3}{2\pi}\Big(X^2-Z^2\Big)+\frac{\kappa_4}{\pi}\Big(X^3-Z^3\Big)=0.
\end{eqnarray}
In the above formulas,  the Taylor-expansion of $\beta\omega$ in $\Phi,\Psi$ was truncated at the fourth-order terms, $\beta\mu_+^D$ and $\beta\mu_-^D$ are given in (\ref{bmu+D}) and (\ref{bmu-D}) respectively, and $X,Z$ are defined in (\ref{XZ}).
Solutions of (\ref{Dbmu+})-(\ref{A-}) determine $\bar c,\bar\zeta,\Phi,\Psi$ for given $T^*,\mu_+,\mu_-$.
\subsection{One-component SALR system}
In the one-component SALR system, the amplitude of the volume-fraction oscillations and the average volume fraction satisfy at the minimum of $\beta\omega$ the following  equations in our MF $\varphi^4$ theory
\begin{equation}
\label{ampl}
\frac{\kappa_4}{3!}\Bigg(\frac{12}{\pi\bar\zeta^3}+A_4(\bar\zeta)\Bigg)\Phi^2-\frac{\kappa_3}{2}\Bigg(\frac{6}{\pi\bar\zeta^2}-A_3(\bar\zeta)\Bigg)\Phi+\frac{6}{\pi\bar\zeta}+A_2(\bar\zeta)-\frac{1}{2T^*}=0
\end{equation}
and 
\begin{eqnarray}
\beta\bar\mu_1&=&\frac{6}{\pi}\Bigg(\ln\Big(\frac{6}{\pi}\bar\zeta\Big)+1\Bigg)+A_1(\bar\zeta)+\beta\tilde V(0)\bar \zeta
\\
\nonumber
 &-&\Bigg(\frac{6}{\pi\bar\zeta^2}-A_3(\bar\zeta)\Bigg)\frac{\Phi^2}{2}+\frac{\kappa_3}{3!}\Bigg(\frac{12}{\pi\bar\zeta^3}+A_4(\bar\zeta)\Bigg)\Phi^3-\frac{\kappa_4}{4!}\Bigg(\frac{9}{\pi\bar\zeta^4}-A_5(\bar\zeta)\Bigg)\Phi^4.
\end{eqnarray}

 \end{document}